\documentclass[twocolumn,twocolappendix]{aastex631}

\usepackage{amsmath}

\newcommand{\figsize}{0.45}
\newcommand{\figsizee}{0.49}
\newcommand{\figsizeee}{0.32}
\newcommand{\xmm}{{\it XMM-Newton}}
\newcommand{\swift}{{\it Swift}-XRT}
\newcommand{\nustar}{{\it NuSTAR}}

\begin{document}

\title{Comparison of X-ray Emission Properties of TDEs and Soft Flares from AGNs}

\correspondingauthor{Samaresh Mondal}
\email{samaresh.astro@gmail.com}

\author[0000-0001-9042-965X]{Samaresh Mondal}
\affiliation{Department of Astronomy, University of Illinois, 1002 W. Green St., Urbana, IL 61801, USA}
\author[0000-0002-4235-7337]{K. Decker French}
\affiliation{Department of Astronomy, University of Illinois, 1002 W. Green St., Urbana, IL 61801, USA}
\author[0000-0001-9668-2920]{Jason T. Hinkle}
\affiliation{Department of Astronomy, University of Illinois, 1002 W. Green St., Urbana, IL 61801, USA}
\affiliation{NSF-Simons AI Institute for the Sky (SkAI), 172 E. Chestnut St., Chicago, IL 60611, USA}
\affiliation{NHFP Einstein Fellow}
%\collaboration{20}{(AAS Journals Data Editors)}

%\author{F.X Timmes}
%\affiliation{Arizona State University}
%\affiliation{AAS Journals Associate Editor-in-Chief}



\begin{abstract}
          X-ray flaring activity from active galactic nuclei (AGNs) may mimic the expected emission from tidal disruption events (TDEs), thus contaminating TDE searches. To compare X-ray emission properties between TDEs and AGNs, we cross-match publicly available \xmm\ and \swift\ point source catalogs with the Million Quasars Catalog and optically selected TDEs. We find that AGNs tend to become softer with a similar hardness ratio (HR) value as TDEs during their flaring events. The soft HRs during flaring events in these AGNs are driven by the emergence of a blackbody-like component below $\sim$2 keV, likely associated with the soft X-ray excess, as well as the continuum emission becoming steeper with the increase of accretion rate. We find 2.5\% in \xmm\ (23 out of 920) and 4.4\% in \swift\ (179 out of 4089) AGNs display flares with peak count rate $>$$2\sigma$ from the median count rate and peak HR$<$$-0.75$. The rate of such flares in \xmm\ and \swift\ is (1.1--2.5)$\times10^{-3}\rm\ galaxy^{-1}\ yr^{-1}$ and (0.9--2.0)$\times10^{-3}\rm\ galaxy^{-1}\ yr^{-1}$, respectively. We also estimate the rate of flares with a maximum flux change (peak/min) of $>20\times$ over on a rest-frame time scale of two years or more with a HR$<$$-0.75$ at peak to be (4.7--11.0)$\times10^{-5}\rm\ galaxy^{-1}\ yr^{-1}$ and (3.1--7.1)$\times10^{-5}\rm\ galaxy^{-1}\ yr^{-1}$ in \xmm\ and \swift, respectively. Finding an optical or infrared counterpart may help to identify these flaring AGNs. We confirm 61\% and 79\% of flaring sources from \xmm\ and \swift, respectively, as AGNs through variability in ZTF light curves or a \emph{WISE} $W1$$-$$W2$$>$0.8 mag color cut.
\end{abstract}



\section{Introduction} \label{sec:intro}
Tidal disruption events (TDEs) are rare phenomena that occur when a star passes too close to a super-massive black hole (SMBH). Nearly half of the disrupted star becomes unbound, and the rest is accreted onto the SMBH via the formation of an accretion disk \citep{rees1988}. The luminosity of TDEs shows a sharp rise and follows a power-law decay over time \citep{rees1988}. TDE candidates were first discovered in the X-ray band by the ROSAT satellite \citep{bade1996,komossa1999b,komossa1999a,grupe1999,greiner2000}. Typically, the X-ray spectra from TDEs are well characterized by a hump-like spectral shape tracing the Wien tail of thermal blackbody emission from the accretion disk. Most TDEs are expected to happen at the lower end of SMBH mass distribution simply because for a sun-like star, the tidal radius will be inside the event horizon radius of $>10^8\ M_{\odot}$ SMBH, and in such a case, no flare will be observed \citep{hills1975}. TDEs are extremely luminous during the peak, and their luminosity can reach near or above the Eddington limit. Therefore, TDEs provide a way to probe the lower end of the SMBH mass distribution at large redshift due to their extreme brightness. However, to probe the SMBH demographics in an unbiased way through TDEs requires a large sample of clean TDEs. The current method for identifying X-ray TDEs relies on a power-law-like decline in the light curve and soft thermal spectrum \citep{khabibullin2014,szanov2021,eyles-ferris2025,grotova2025}. However, similar properties can be seen in other extragalactic transient events such as X-ray flares from active galaxies, which are thought to be the prime contaminants in X-ray TDE searches.

Active galaxies, also known as active galactic nuclei (AGNs), which harbor a long-lived accretion disk, emit across multiple wavebands, including X-rays. The variability in AGN X-ray light curves is a common feature that spans several orders of magnitude in brightness. The source of large-scale X-ray variability in AGN is not clear yet and can have multiple origins, such as changes in accretion rate, magnetic reconnection in the hot corona leading to flares, or obscuration due to intervening clouds in the dusty torus passing through the line of sight. Typically, X-ray emission from AGN is harder than that from TDEs. In a $z\sim1$--2 universe, AGN spectra are best fitted by a power-law index $\Gamma\sim1.75$--1.95 \citep{tozzi2006,liu2016,mareschi2016} indicating a non-thermal origin of the X-ray emission. 

However, AGN variability on month-years timescales can potentially be misclassified as coming from a TDE. AGN have been observed to exhibit softer spectra during their flare states \citep{sobolewska2009}, which is a similar signature to that expected for a TDE flare. Several searches for large X-ray flares have used flare amplitude, X-ray spectra, and multi-wavelength properties to look for separation between AGN and TDEs. For example, \citet{esquej2007} compared the ROSAT and \xmm\ slew catalogs and found five sources with large-scale flux variation by a factor $>20$. Further follow-up with optical spectra indicates that two were likely to be due to AGN variability, two were good candidates for TDEs, while XMMSL1 J024916.6--041244 was apparently a persistent Seyfert 1.9 galaxy but showed TDE-like traits such as a very soft X-ray spectrum. More recently, \citet{auchettl2018} compared a sample of highly variable AGN and four TDEs, finding that TDE light curves follow a monotonic decline with a power-law index ranging from 0 to 2. On the other hand, the power-law index for the AGN flare light curves can range from $-10$ to $+15$, depending on the time scale, highlighting a diversity in AGN flare characteristics. Furthermore, \citet{auchettl2018} found that during the flare, AGN spectra tend to become softer with similar hardness ratios (HR) values as TDEs. Both the diversity in AGN flares and softer spectra during the flare make AGN flares difficult to separate from TDEs. In contrast, \citet{guolo2024} also compared the X-ray spectra of a sample of optically selected X-ray bright TDEs with the long-term time-averaged X-ray spectra of type I and II AGNs and found that TDEs and AGNs have a clear separation in HR or X-ray spectral photon index $\Gamma$.

The soft X-ray flares from AGN can be distinguished from TDEs with pre-existing optical and infrared (IR) observations. The central part of AGN is surrounded by a dusty torus, and a considerable fraction of high-energy emission is absorbed by dust, which then re-radiates in IR bands. Previous studies have shown IR color can be highly effective in identifying AGNs by cross-matching \emph{WISE} catalog with known AGNs \citep{assef2010,stern2012}. Furthermore, the IR color cut can be utilized to identify highly obscured AGN in which the emission from the accretion disk is blocked by the dusty torus \citep{mateos2013,stern2014}. Typically, a threshold of $W1-W2>0.8$ mag \emph{WISE} color cut can reach a reliability of 95\% to a depth of $W2\sim15$ mag in identifying AGNs in the COSMOS field \citep{stern2012,assef2013}. In addition to the IR color, the AGNs can also be identified through the variability in the optical light curves. The variability in AGN optical light curves is stochastic in nature and can be well fitted by a damped random walk (DRW) model \citep[e.g.,][]{kelly2009,macleod2010,kozlowski2016,burke2021}. A DRW model has only two free parameters, the characteristic time scale and the variability amplitude, and the variability amplitude has been largely effective in identifying AGNs \citep{baldassare2018,baldassare2020}.

In this paper, we compare the X-ray emission properties of TDE with flares from AGN. We search for AGN flares with variation $>2\sigma$ from the median pre-flare count rate and peak $\rm HR<-0.75$ mimicking a TDE. Furthermore, we quantify the rate of such flares from AGN and compare it with the current TDE rate. Distinguishing flares from AGNs and TDEs is a major challenge in time-domain astronomy. AGNs are highly variable, stochastic, and recurring, whereas a typical TDE presents as a single, rapid outburst followed by a monotonic decay over months or years. The goal of this analysis is to identify AGN flares with similar behavior to TDEs and provide a framework for the selection of clean TDEs that can be applied to currently operating all-sky X-ray surveys, such as eROSITA and the Einstein probe.

\begin{figure*}[]
\centering
\includegraphics[width=\figsizee\textwidth]{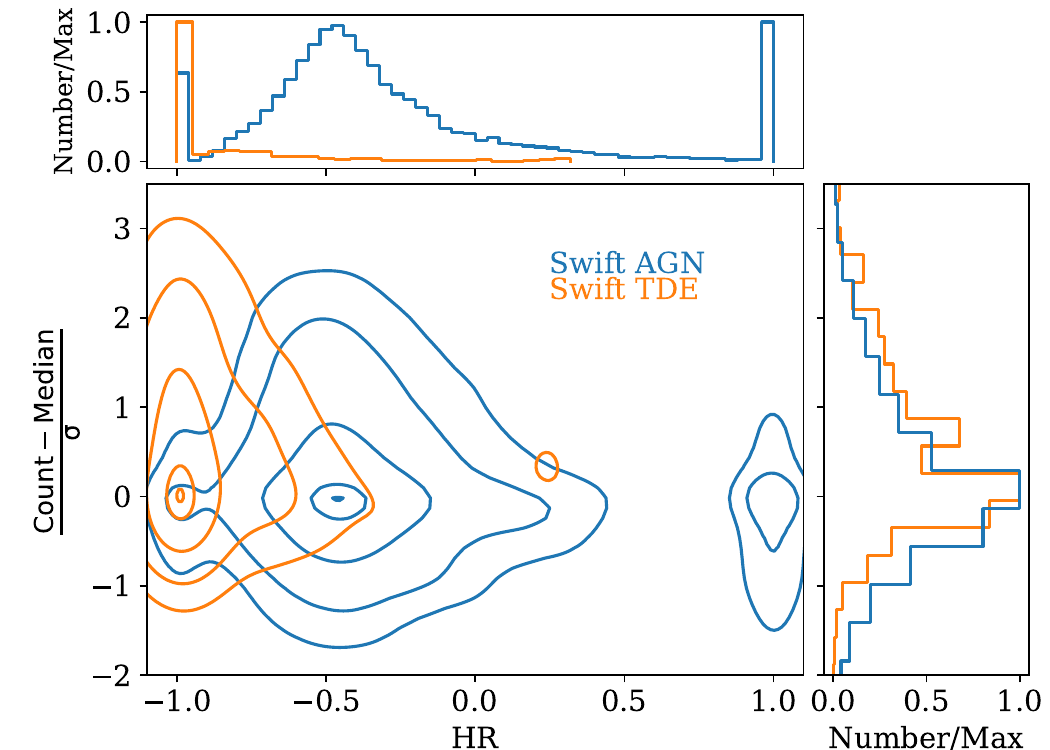}
\includegraphics[width=\figsizee\textwidth]{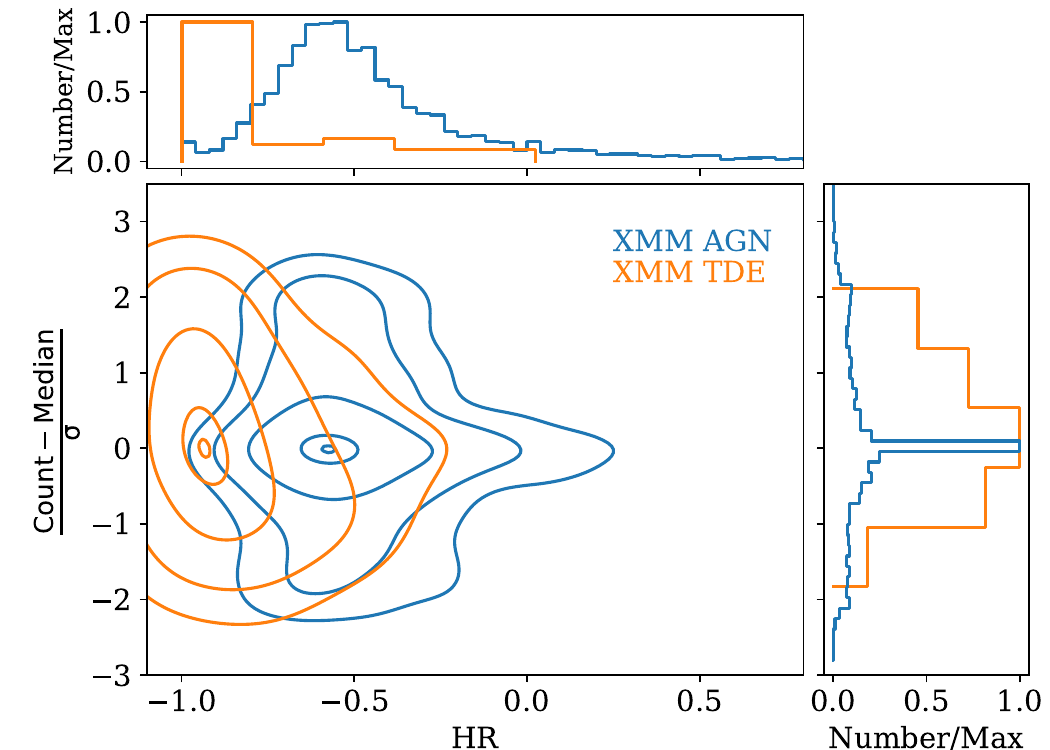}
\includegraphics[width=\figsizee\textwidth]{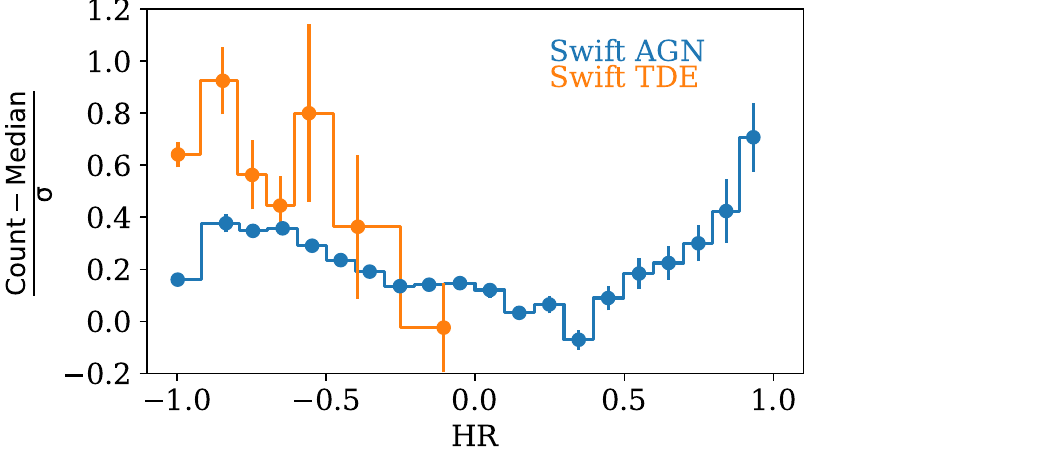}
\includegraphics[width=\figsizee\textwidth]{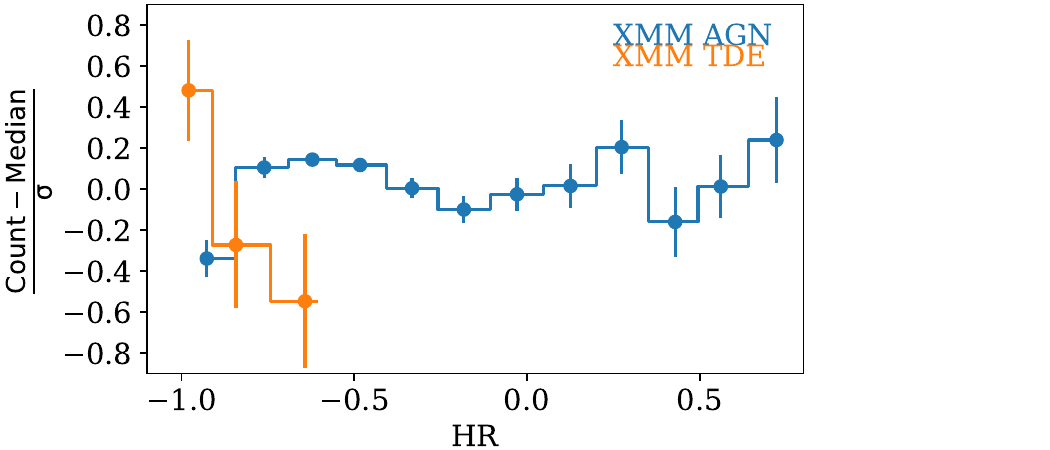}
\caption{Top panels: contour plot (17\%, 34\%, 68\%, 95\%, and 99.7\%) of changes in the count rate from median vs HR for TDEs and AGNs obtained from the \swift\ (left) and \xmm\ (right) data sets. Bottom panels: same as the top panels, but the individual measurements of TDEs and AGNs are binned with a step size of $\rm \Delta HR=0.1$. Both panels indicate there is a significant overlap between AGNs and TDEs.}
\label{fig:hr_agn_tde}
\end{figure*}

\section{Methods}\label{sec:data}
We utilize the publicly available \swift\  \citep{burrows2005} living catalogue LSXPS \citep{evans2023} and \xmm\ \citep{jansen2001} stack catalog 4XMM-DR14s \citep{traulsen2019,traulsen2020} to construct long-term X-ray light curves for point sources. While constructing the long-term light curves, we use various flags to avoid pitfalls that may lead to spurious detection. The details of the flags used here are similar to \S2.1 in \citet{mondal2025}; differences are noted in the following paragraphs. Furthermore, we use the same procedure mentioned in \citet{mondal2025} for cross-matching the X-ray point source catalogs with the Million Quasars Catalog \citep{flesch2023} to identify X-ray detected AGNs and sources specifically classified as TDEs in TNS\footnote{https://www.wis-tns.org}. We calculate the nearest neighbor distance between the X-ray point source catalogs and AGN, TDE catalogs. Furthermore, we use a maximum separation of $6''$ for \xmm\ and $19''$ for \swift\ to identify their X-ray counterparts. We also check if there are two or more X-ray sources within the limiting distance for cross-matching; then, in such a case, we do not consider that source. Such a situation can arise if there are X-ray binaries and/or ultraluminous X-ray sources close to the central AGN. In addition to that, we ensure all the AGNs must have at least three observations, as we are interested in AGNs that have flare-like events.

To identify the flaring AGNs, for the individual sources, we compute $\frac{\rm Count-Median}{\sigma}$, indicating how strong the flare is from the median count rate, where Count is the integrated count rate in a given observation and Median, $\sigma$ are the median and standard deviation of the count rates across multiple observations. For each observation, we also compute the HR from the count rate in the soft 0.3--2 keV and hard 2--10 keV bands as $\rm HR=\frac{C_{hard}-C_{soft}}{C_{hard}+C_{soft}}$. $\rm HR=-1$ and $+1$ correspond to completely thermal and non-thermal emission, respectively. We note that in the \swift\ AGN sample, around 23\% of the observations have $\rm HR=-1$ compared to 1\% in \xmm. This is primarily because, due to low exposure in \swift, many of the AGNs are marginally detected, only in the soft band, which leads to an HR value of $-1$. Therefore, in the case of \swift\ observations, we require that if $\rm HR=-1$ then the count rate in the soft band should be $\rm C_{\rm soft}>3\times10^{-2}\ count\ s^{-1}\ cm^{-2}$ which is around 10 times higher than the detection sensitivity limit in 0.3--10 keV band for a 2 ks exposure. This ensures that all the $\rm HR=-1$ observations in \swift\ sample are due to intrinsically soft spectra. After doing this filtering, the percentage of \swift\ AGN observations with $\rm HR=-1$ is reduced to 4.6\% of total observations. We find 920 AGNs in \xmm\ and 4089 AGNs in \swift\ with at least three observations that satisfied our above-mentioned criteria. AGNs may show intra-observational flares with a time scale of a few seconds to an hour. However, these flares will be missed as we are integrating the counts over the entire observation. This will have a very little effect on our AGN and TDE comparison, as we are interested in large-scale AGN flares with time scales of days to multiple weeks mimicking a TDE. The \swift\ TDE sample consists of 51 unique TDEs with a total number of 1450 observations. The median number of observations per TDE is 16, with a standard deviation of 34. The \xmm\ TDE sample consists of 13 unique TDEs with a total number of 42 observations, which have a median number of three observations per TDE. The details of each TDE are given in Table \ref{tab:tde_list}. While constructing our TDE sample, we do not include Swift J1644+57, Swift J2058+05, and AT2022cmc. These are three well-known jetted TDEs that have a very hard X-ray spectrum, possibly originating from a jet, and do not represent the vast majority of the TDEs where the X-ray emission likely comes from an accretion disk.

\begin{figure}[]
\centering
\includegraphics[width=\figsize\textwidth]{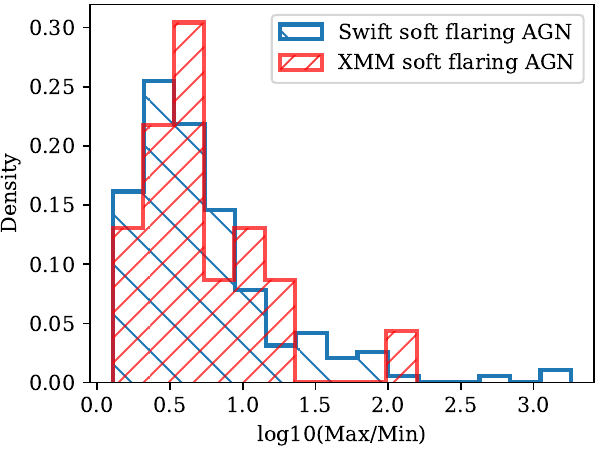}
\caption{Variability range of soft flaring AGNs selected using the cut $\frac{\rm Count-Median}{\sigma}>2$ and $\rm HR<-0.75$. Among the $2\sigma$ flares we select from \swift\ and \xmm\, 50\% have a ratio of maximum to minimum flux $>3.9$ and $>4.5$, respectively.}
\label{fig:maxmin_var}
\end{figure}

\section{Results}\label{sec:results}
\subsection{X-ray HR comparison of AGN flares and TDEs}
Figure \ref{fig:hr_agn_tde} (top panels) shows the distribution of HR and $\frac{\rm Count-Median}{\sigma}$ for the AGN and TDE samples. There is a significant overlap between the two populations of sources. The majority of the AGN detections are gathered around $\rm HR\sim-0.5$ in both \swift\ and \xmm\ data sets. The TDE sample is clustered near $\rm HR\sim-1$ with an extended tail towards positive HR values that creates a significant overlap between the TDEs and AGNs.

The contour plot in Figure \ref{fig:hr_agn_tde} shows a trend of spectral softening as the count rate increases from the median value. To quantify this trend, we bin the data with a step size of $\rm \Delta HR=0.1$. The bottom panels of Figure \ref{fig:hr_agn_tde} show the binned data, and the error bars indicate the standard error of the mean in each bin. For the \swift\ AGN sample with $\rm HR<+0.3$, the binned data show the count rate increases from the median as HR softens. A similar trend of $\frac{\rm Count-Median}{\sigma}$ increase with the softening of HR is also visible for the TDE sample.

In the \xmm\ data set of the AGN sample, we see a similar trend of becoming softer as deviating from the median value, but less pronounced compared to the \swift\ data set. In the AGN sample, we see a reversal at $\rm HR\gtrsim+0.3$ where the sources tend to become harder for the large-scale variations. These might be associated with obscured type II Compton-thick AGNs in which the HR concentration at $+1.0$ is driven by large absorption column density intrinsic to the sources. To reveal the exact nature of these sources that show a harder-when-brighter trend requires a detailed investigation, which is beyond the scope of this paper. We are interested in AGNs that have a flare but a very soft X-ray spectrum that might mimic a TDE. The median HR value (with $1\sigma$ standard deviation) for the detections of TDEs is $-1.0(\pm0.28)$ and $-0.92(\pm0.27)$ in \swift\ and \xmm\ sample, respectively. Therefore, for the subsequent analysis, we choose $\rm HR<-0.75$ (around median+$1\sigma$ between \swift\ TDE sample) to identify soft AGN flares that have TDE-like spectral traits and large-scale variation with $\frac{\rm Count-Median}{\sigma}>2$. In \swift\ and \xmm\ we find 179 and 23 such AGNs, respectively.

Our selection of flares with $\frac{\rm Count-Median}{\sigma}>2$ produces a range of flare magnitudes, as $\sigma$ depends on the source variability and measurement uncertainties. In Figure \ref{fig:maxmin_var}, we show the variability range of the selected AGNs with the cut $\frac{\rm Count-Median}{\sigma}>2$ and $\rm HR<-0.75$. Among the $2\sigma$ flares we select from \swift\ and \xmm\, 50\% have a ratio of maximum to minimum flux $>3.9$ and $>4.5$, respectively. Thus, these flares are largely comparable with the overall magnitude of flares that exhibit high maximum variability from \citet{mondal2025}.

\begin{figure*}[]
\centering
\includegraphics[width=\figsizeee\textwidth]{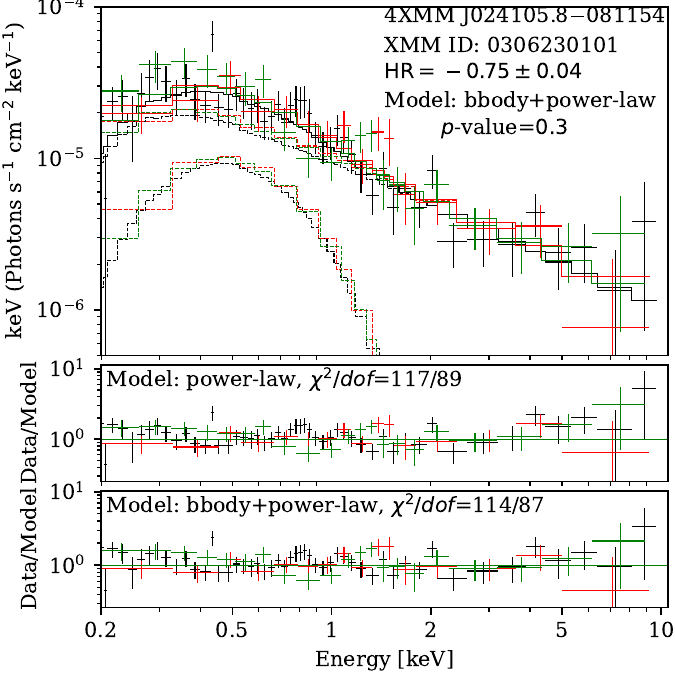}
\includegraphics[width=\figsizeee\textwidth]{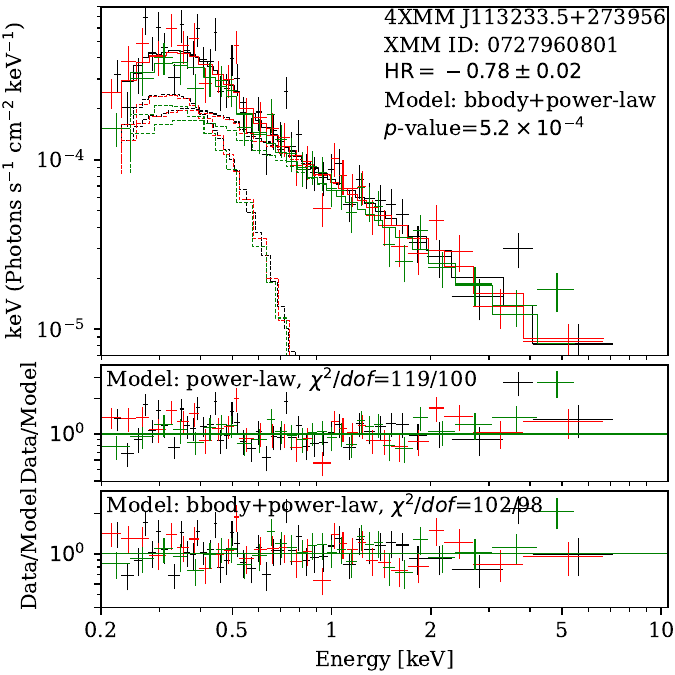}
\includegraphics[width=\figsizeee\textwidth]{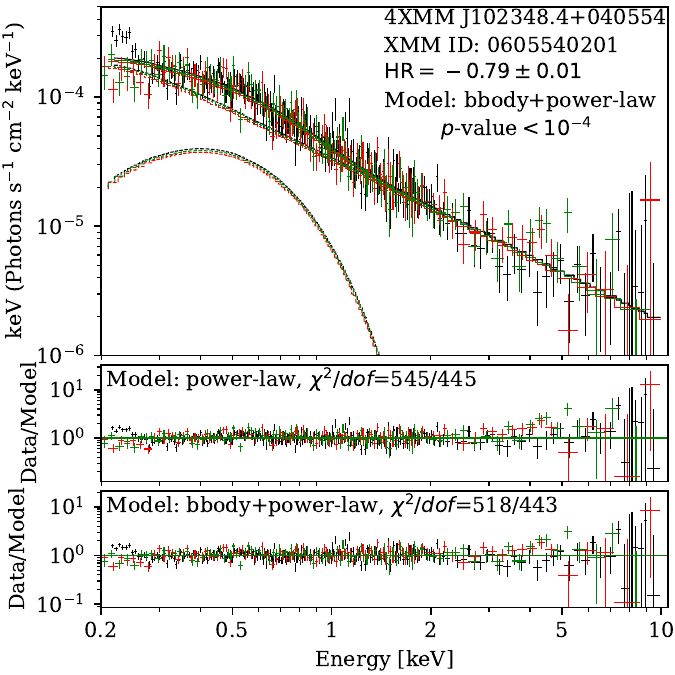}
\includegraphics[width=\figsizeee\textwidth]{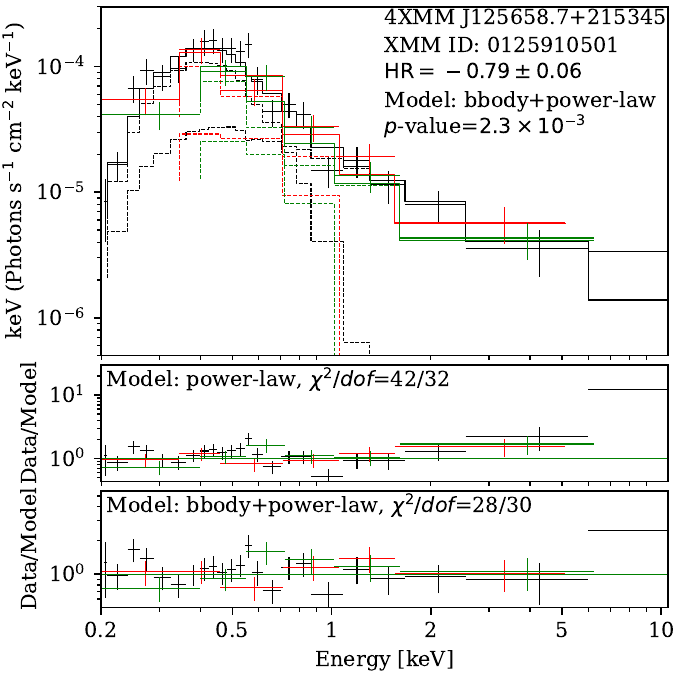}
\includegraphics[width=\figsizeee\textwidth]{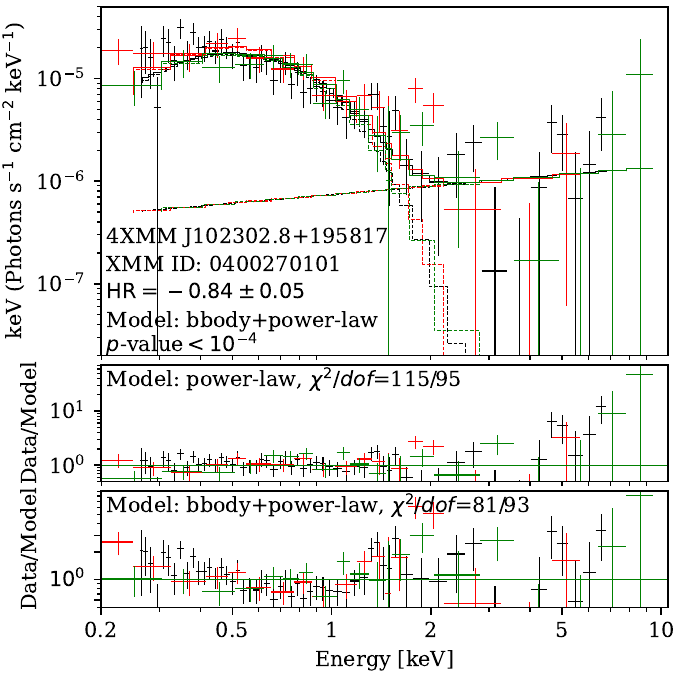}
\includegraphics[width=\figsizeee\textwidth]{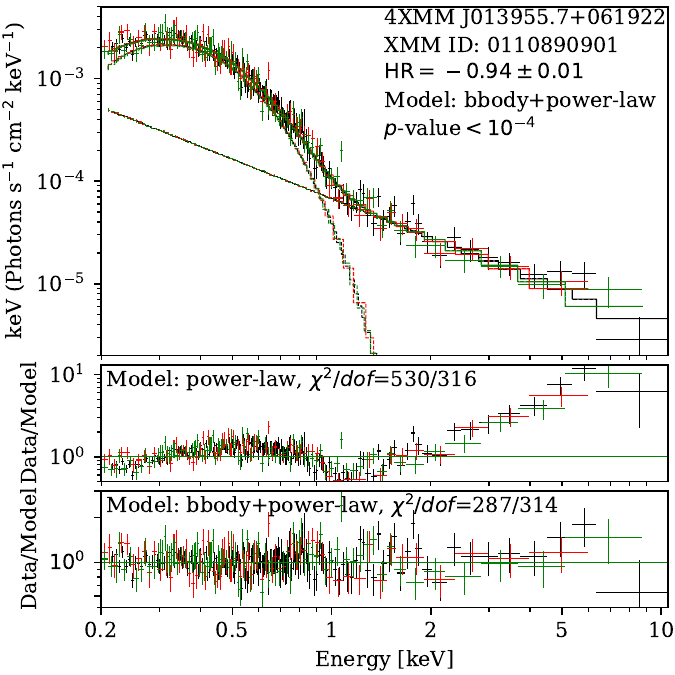}
\caption{\xmm\ spectra of six bright soft flaring AGNs with $\rm HR<-0.75$ and $\frac{\rm Count-Median}{\sigma}>2$ that have net count above 1000. The $p$-value in each inset indicates the statistical improvement in the fit after adding a blackbody component compared to a single power-law model. Except for the source 4XMM J024105.8--081154, the blackbody component is required to model the broad-band X-ray spectrum. Furthermore, the strength of the blackbody component over the power-law component increases for lower values of HR. The plotted spectra are in the observer frame. The black, red, and green colors indicate data points from EPIC-pn, MOS1, and MOS2 detectors, respectively.}
\label{fig:xmm_src_fit}
\end{figure*}

\subsection{X-ray spectra of soft AGNs}

\subsubsection{\xmm\ individual spectra of soft AGNs}
\begin{figure*}[]
\centering
\includegraphics[width=\figsizeee\textwidth]{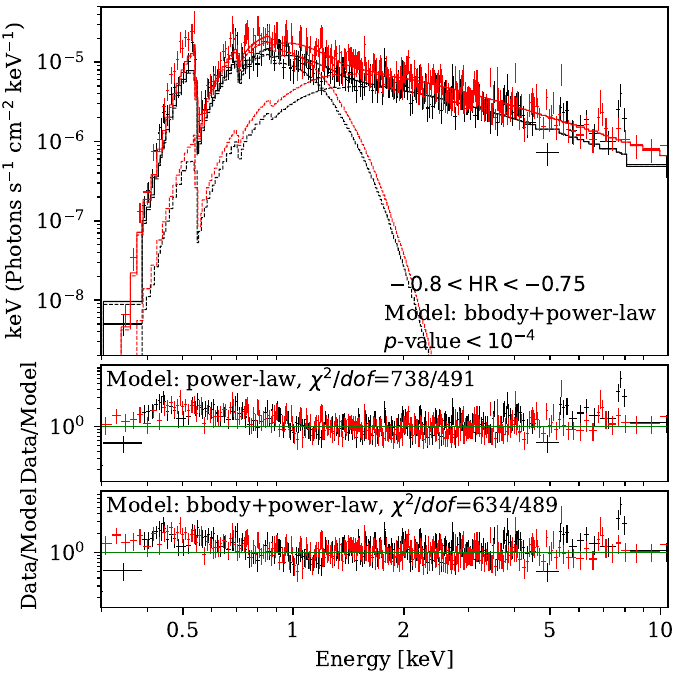}
\includegraphics[width=\figsizeee\textwidth]{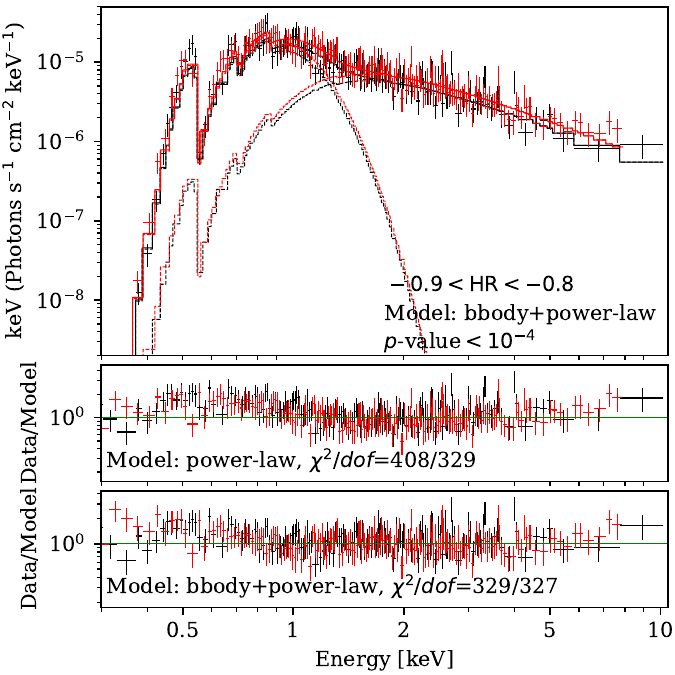}
\includegraphics[width=\figsizeee\textwidth]{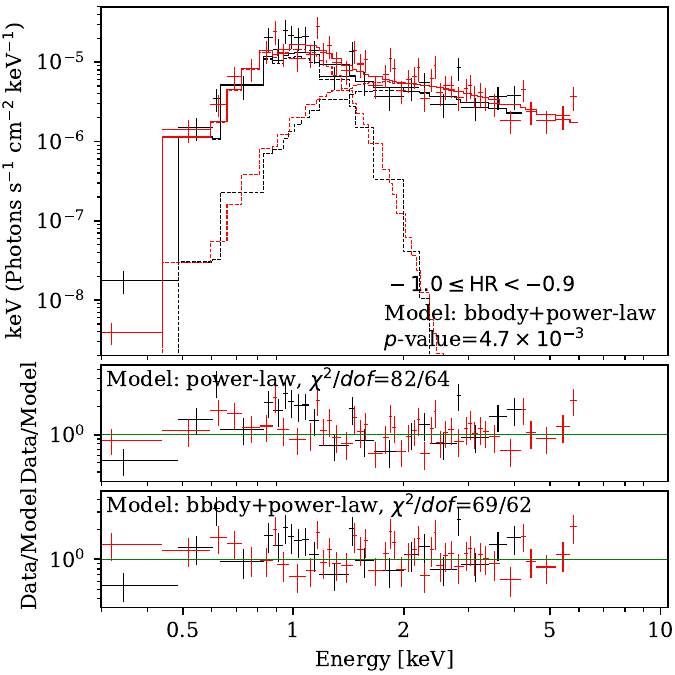}
\caption{\xmm\ stack spectra of all faint AGNs with net count less than 1000 for different HR groups. A trend similar to Figure \ref{fig:xmm_src_fit} is also visible, where adding a blackbody component over a power-law component greatly improves the overall fit. Black and red represent stacked data points from EPIC-pn and MOS1/MOS2, respectively. The unfolded spectra of stack sources peak at 0.5--1 keV compared to the individual spectra, which typically peak below 0.5 keV, primarily because the stack spectra are plotted in the rest frame, whereas the individual spectra in Figure \ref{fig:xmm_src_fit} are in the observed frame.}
\label{fig:xmm_stack_fit}
\end{figure*}
The X-ray spectra may provide a clue for which spectral component is responsible for the extreme softness in these flaring AGNs. Among the 23 \xmm\ soft flaring AGNs, only six sources have 0.3--10 keV band net count above 1000: 4XMM J024105.8--081154 (ObsID: 0306230101), 4XMM J113233.5+273956 (ObsID: 0727960801), 4XMM J102348.4+040554 (ObsID: 0605540201), 4XMM J125658.7+215345 (ObsID: 0125910501), 4XMM J102302.8+195817 (ObsID: 0400270101), and 4XMM J013955.7+061922 (ObsID: 0110890901). We extracted the \xmm\ spectra of these six AGNs, and for the X-ray spectral fitting, we use the software XSPEC \citep{arnaud1996}. The spectra of these six observations correspond to the epochs with a $2\sigma$ flare above the median count rate and $\rm HR<-0.75$. For the line of sight absorption column density, we used two absorption components with the first one \texttt{tbabs} \citep{wilms2000} fixed to the Galactic line of sight column density obtained from HI4PI \citep{hi4pi2016} and the other one \texttt{ztbabs} kept free for the intrinsic absorption associated with the source. First, we tried to fit the spectra of the six sources using an absorbed power-law model. However, we notice that for all sources, adding a blackbody component improves the fit. Furthermore, we performed an F-test of $\chi^2$ improvement of the fit and find in all cases the blackbody plus power-law model is favored over power-law model with a statistical significance of $p$-value$\ \leq2.3\times10^{-3}$ except for the case of 4XMM J024105.8--081154, indicating a single power-law model can be rejected at more than $3\sigma$ confidence level. For the source 4XMM J024105.8--081154, adding a blackbody component does not improve the fit significantly, only by $\Delta\chi^2=3$ for two additional degrees of freedom with $p$-value$\ =0.3$. Figure \ref{fig:xmm_src_fit} shows the unfolded \xmm\ spectra of these six bright sources that are fitted with an absorbed blackbody plus power-law model. Figure \ref{fig:xmm_src_fit} also shows that the contribution from the blackbody component to the total spectrum increases for lower values of HR, which suggests that for a lower HR value, the higher contribution from the blackbody-like component at energies 0.3--2 keV can make the overall spectrum much softer.

In addition to the above analysis, we fitted the spectra of all bright soft AGN observations (no $\frac{\rm Count-Median}{\sigma}$ cut) with $\rm HR<-0.75$ and net count above 1000 by a two-component model composed of an absorbed blackbody plus power-law and compared the fit statistics with a simple power-law model. In the \xmm\ AGN sample, there are 46 observations with $\rm HR<-0.75$ and net count above 1000. Of these 46 observations (of 13 unique AGNs), 32/46 (70\%) favor a blackbody plus power-law model at $>2\sigma$ confidence level over a simple power-law model. Considering the subset of 23 observations with $\rm HR<-0.8$, 18/23 (78\%) favor a blackbody plus power-law model.

\begin{figure*}[]
\centering
\includegraphics[width=\figsizeee\textwidth]{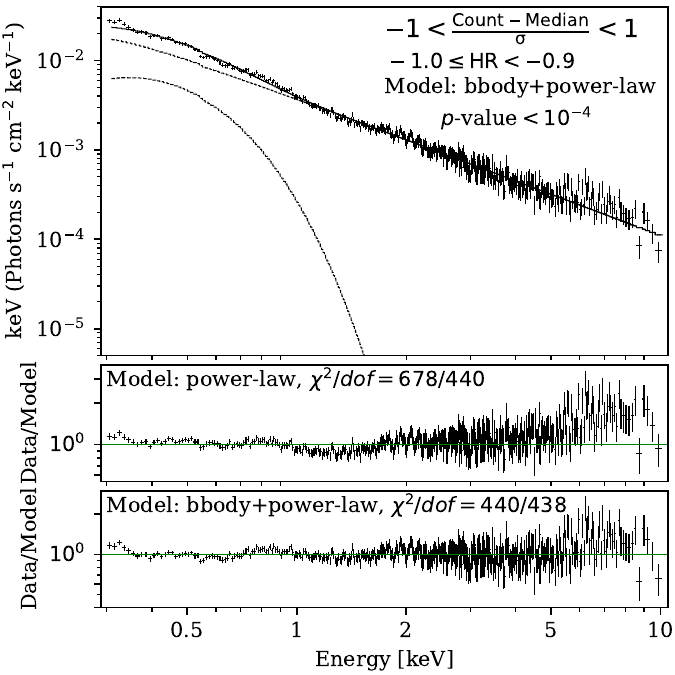}
\includegraphics[width=\figsizeee\textwidth]{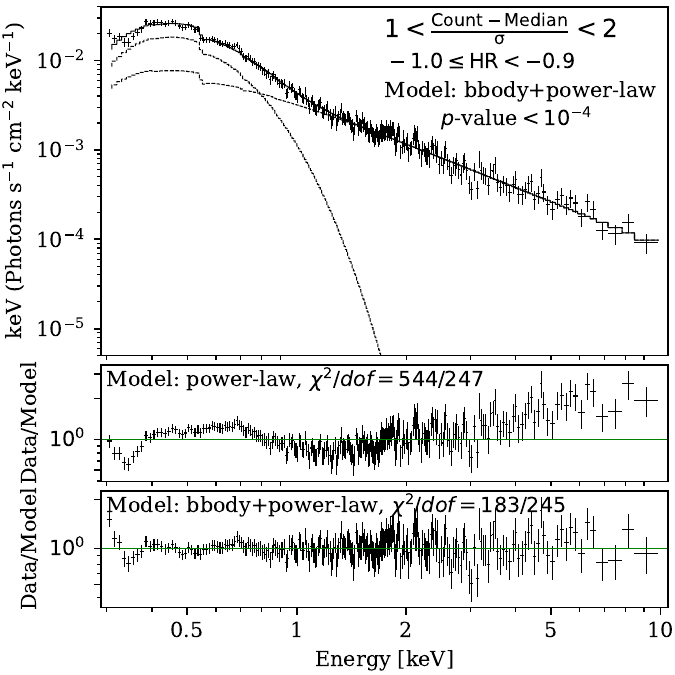}
\includegraphics[width=\figsizeee\textwidth]{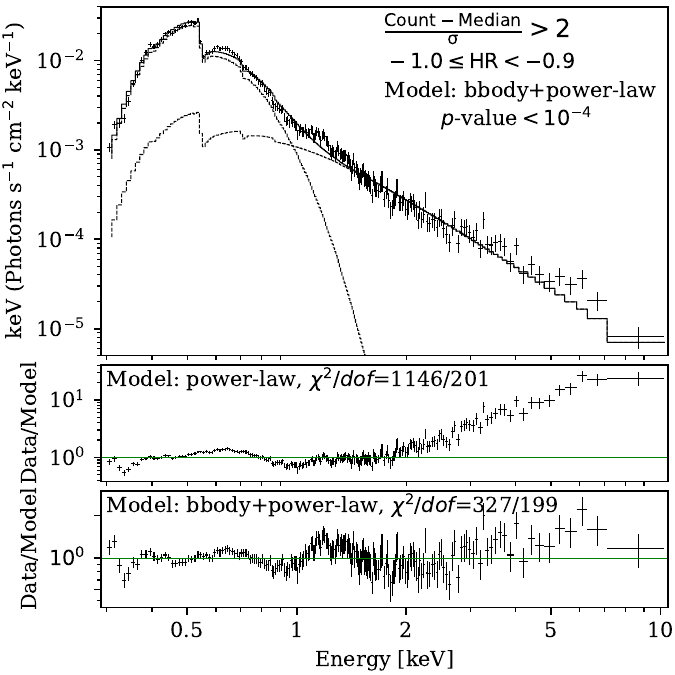}

\includegraphics[width=\figsizeee\textwidth]{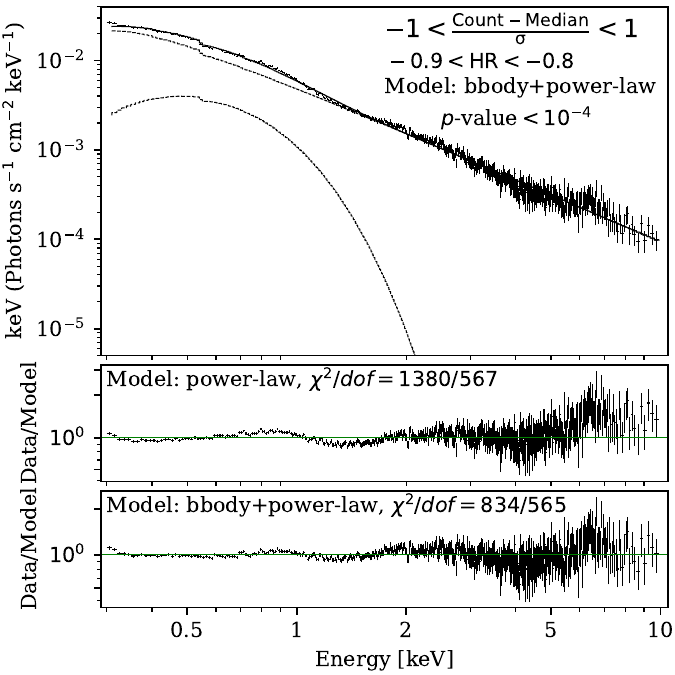}
\includegraphics[width=\figsizeee\textwidth]{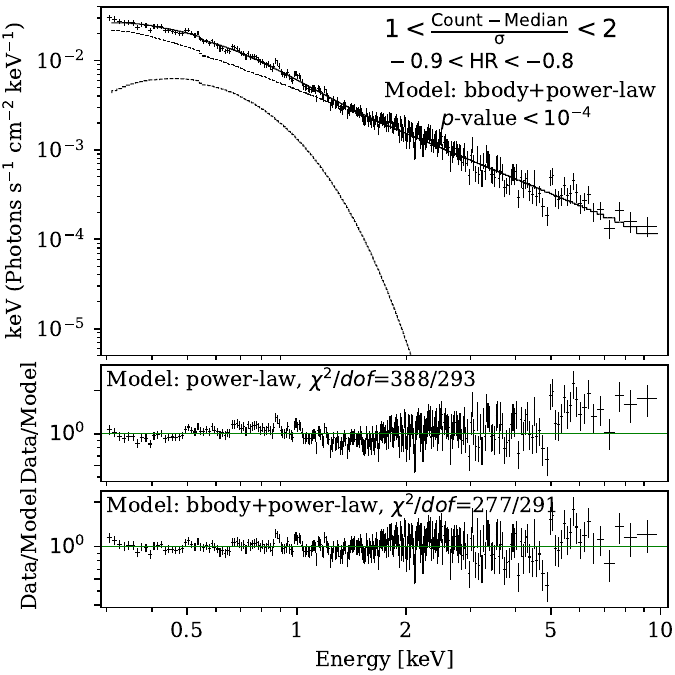}
\includegraphics[width=\figsizeee\textwidth]{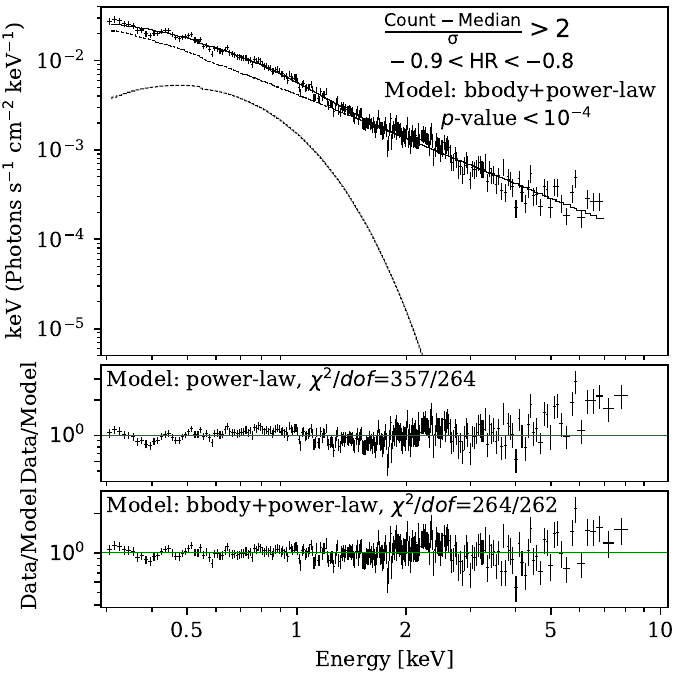}

\includegraphics[width=\figsizeee\textwidth]{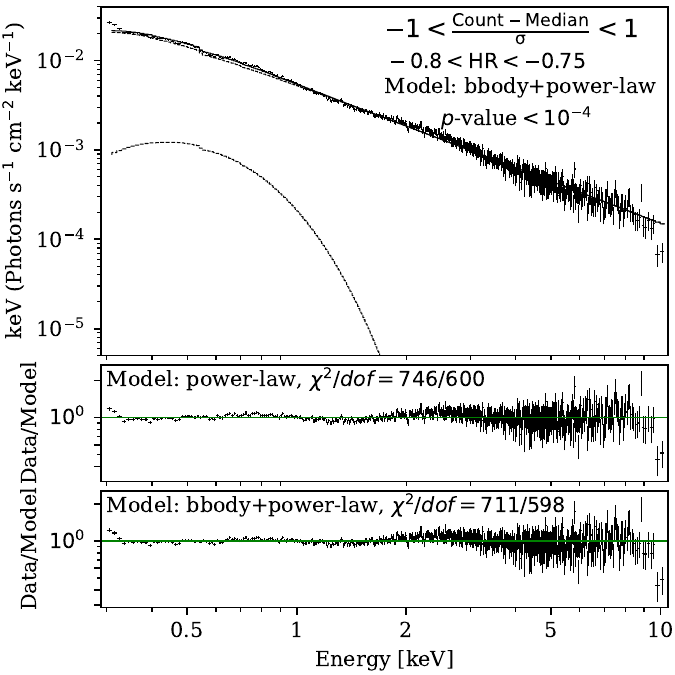}
\includegraphics[width=\figsizeee\textwidth]{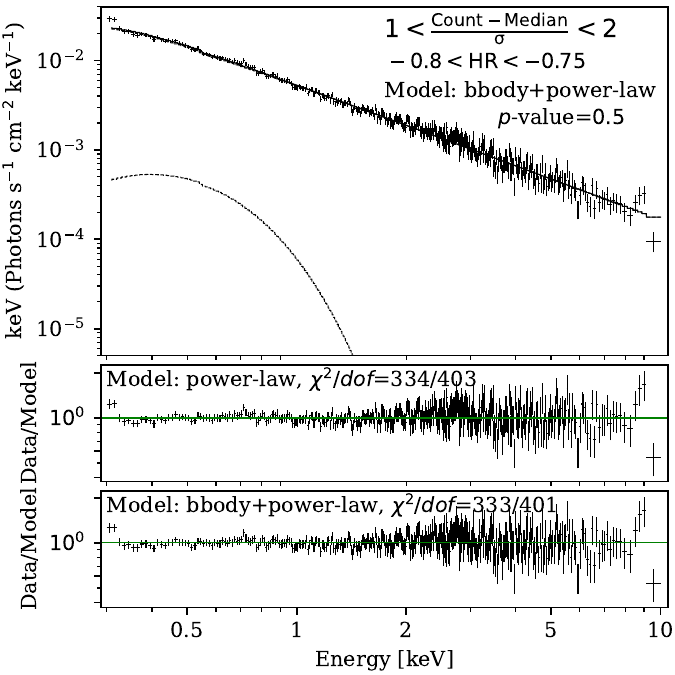}
\includegraphics[width=\figsizeee\textwidth]{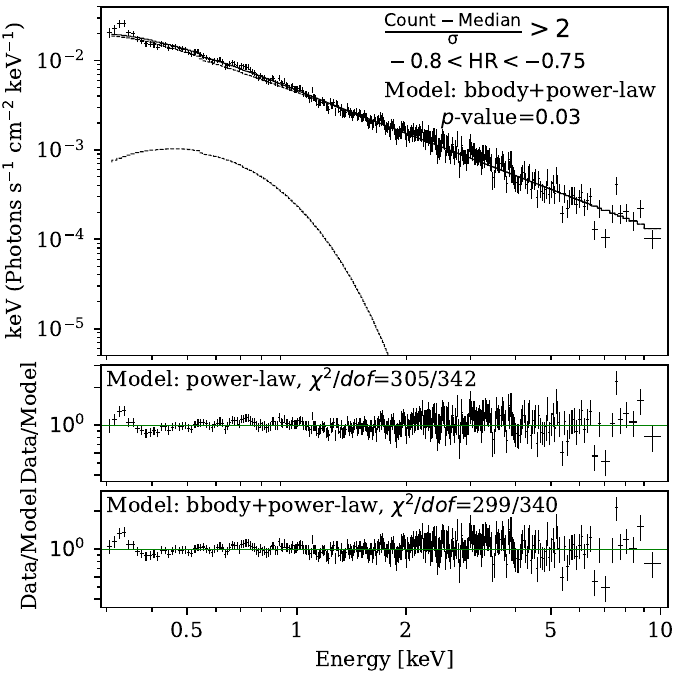}
\caption{\swift\ stack spectra of all soft AGNs for different HR and $\frac{\rm Count-Median}{\sigma}$ groups. Stacks are shown with increasing flare strength from left to right, and increasing HR from bottom to top. A similar trend as in the \xmm\ data is seen here. In general, a blackbody component improves the fit for stacked spectra with $\rm HR<-0.8$, and the contribution of the blackbody component increases for higher values $\frac{\rm Count-Median}{\sigma}$. The unfolded spectra and model components are rescaled for plotting purposes to have a similar y-axis range across different groups.}
\label{fig:swift_stack_fit}
\end{figure*}

\subsubsection{\xmm\ stacked spectra of soft AGNs}
The above spectral modeling of individual \xmm\ AGNs may not represent the overall sample of the soft AGNs, as the number of \xmm\ observations with over 1000 total counts and $\rm HR<-0.75$ represents only $\sim11\%$ (46 out of 427) of all soft AGN observations with $\rm HR<-0.75$. Therefore, to assess the spectral decomposition of the overall population of the soft AGNs, we stacked the spectra of soft AGNs with $\rm HR<-0.75$ and net count less than 1000. We stacked the spectra of soft AGNs into three HR groups: $\rm -0.8<HR<-0.75$, $\rm -0.9<HR<-0.8$, and $\rm -1.0\leq HR<-0.9$. We utilized the software \texttt{Xstack}\footnote{https://github.com/AstroChensj/Xstack}\footnote{https://doi.org/10.5281/zenodo.20327926} \citep{chen2025} that shifts the spectra of all sources to the common rest-frame for a given redshift before performing the stacking. Different sources have varying Galactic line of sight column density $N_{\rm H,Gal}$. Therefore, for the correction of the Galactic line of sight absorption column density, before doing the stacking, the ancillary response file of each source is multiplied by the foreground absorption curve for a given $N_{\rm H,Gal}$ value. The responses (ARF$\times$RMF) are stacked with proper weights that optimally preserve intrinsic spectral shape \citep[see Appendix A of][]{chen2025}. While doing the spectral fitting of stacked spectra, we include only one absorption component for the intrinsic absorption associated with the sources, as the correction for the Galactic line of sight absorption is already applied. Figure \ref{fig:xmm_stack_fit} shows the stacked spectra of the low-count soft AGNs. The blackbody component is required for all three HR groups with extremely high significance ($p$-value$\ \leq4.7\times10^{-3}$), similar to the spectral analysis of individual \xmm\ flaring AGNs.

\subsubsection{\swift\ stack spectra of soft AGNs}
The typical exposure of \swift\ observation is below $\sim$2 ks, and \swift\ effective area is 10 times smaller than \xmm\ EPIC-pn. Therefore, individual \swift\ observations do not have high enough counts for spectral modeling of the soft flaring AGNs. Therefore, we performed stacked spectral modeling for three HR groups similar to the \xmm\ faint soft AGN sample. However, the number of observations of soft flaring AGN with $\rm HR<-0.75$ in \swift\ is much larger than the \xmm\ sample (4010 in \swift\ compared to 427 in \xmm); therefore it is possible to test the dependence of spectral components with variability in stacked spectra by dividing the soft AGN sample into three groups: $-1<\frac{\rm Count-Median}{\sigma}<1$, $1<\frac{\rm Count-Median}{\sigma}<2$, and $\frac{\rm Count-Median}{\sigma}>2$.

Figure \ref{fig:swift_stack_fit} shows the \swift\ stacked spectra of different groups. A trend similar to \xmm\ data is also visible here. Adding a blackbody for stacked spectra with $\rm HR<-0.8$ provides a significant improvement to the overall fit statistics. For the $\rm -0.8<HR<-0.75$ group, adding a blackbody component over the power law component does not provide much $\Delta\chi^2$ improvement for all three $\frac{\rm Count-Median}{\sigma}$ bins. We find that the strength of the blackbody component over the power-law component increases for higher values $\frac{\rm Count-Median}{\sigma}$ as well as lower values of HR, indicating a spectral variation not only based on the HR but also the magnitude of flares. The \swift\ stack spectral analysis suggests that the larger the AGN flare is, the softer its spectrum will become.  

\subsection{Optical and IR counterparts}
\label{sec:counterparts}
We search for optical and IR counterparts of soft flaring AGNs with $\rm HR<-0.75$ and $\frac{\rm Count-Median}{\sigma}>2$. Finding optical and IR counterparts may help identify these soft flaring sources as AGNs through the stochastic variability in their light curves and IR colors, and avoid confusion with TDEs. The goal of the optical IR counterpart search is to check how many flaring sources selected from X-rays without any prior classification can be identified as AGNs.

\begin{figure}[]
\centering
\includegraphics[width=\figsize\textwidth]{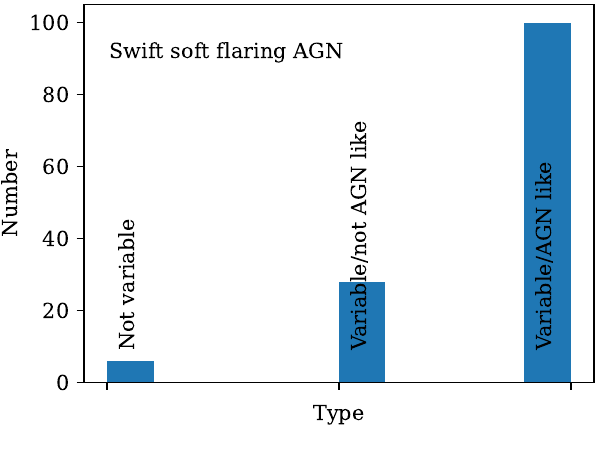}
\includegraphics[width=\figsize\textwidth]{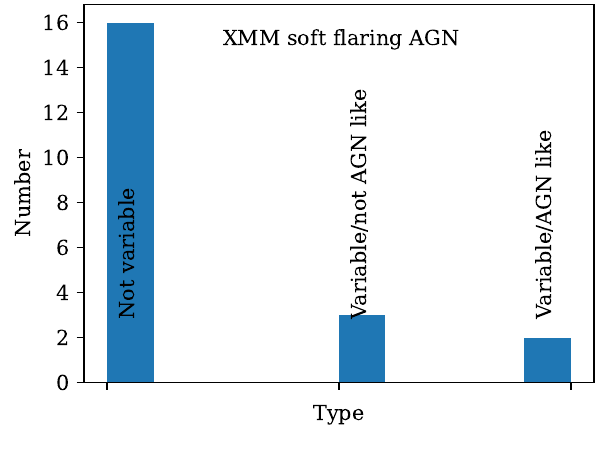}
\caption{Cross-match between soft flaring AGNs with ZTF, and the identifications are based on variability in the optical light curve.}
\label{fig:xray_ztf}
\end{figure}

We search for optical counterparts in ZTF \citep{bellm2019}. Among the 23 \xmm\ and 179 \swift\ soft flaring AGNs, we find ZTF counterparts for 21 and 134 sources, respectively. AGNs exhibit variability across multiple wavebands on a timescale ranging from hours to decades. AGN variability is stochastic in nature with flux differences that tend to be larger on larger timescales. The optical variability in AGN light curves is best described by a DRW model \citep{kelly2009,macleod2010,kozlowski2010,burke2021}. Therefore, the optical variability can be used to identify a potential AGN by searching for variability consistent with a DRW model. We use the QSO fitting software \citep{butler2011} that employs a DRW model and returns the best-fit model parameters and significances $\sigma_{\rm vary}$, $\sigma_{\rm QSO}$, and $\sigma_{\rm notQSO}$ with corresponding $\chi_{\rm vary}$, $\chi_{\rm QSO}$, and $\chi_{\rm notQSO}$. These $\sigma$ values represent the significance of source variability, how significant the source variability is DRW-like, and the significance that source variability is random, not AGN-like. Sources with $\sigma_{\rm vary}<3$ are designated as not variable. Sources with $\sigma_{\rm vary}>3$ and $\sigma_{\rm QSO}>\sigma_{\rm notQSO}$ is identified as AGN whereas $\sigma_{\rm vary}>3$ and $\sigma_{\rm QSO}<\sigma_{\rm notQSO}$ defined as variable not AGN like. Figure \ref{fig:xray_ztf} shows the classification of the sources based on ZTF light curves. Among the 21 \xmm\ flaring sources that have a ZTF counterpart, only two can be identified as AGN; on the other hand, among the 134 \swift\ flaring sources that have a ZTF counterpart, 100 can be identified as AGN.

\begin{figure}[]
\centering
\includegraphics[width=\figsize\textwidth]{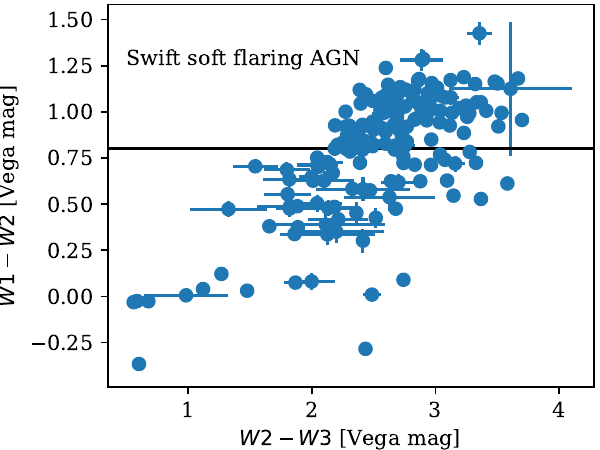}
\includegraphics[width=\figsize\textwidth]{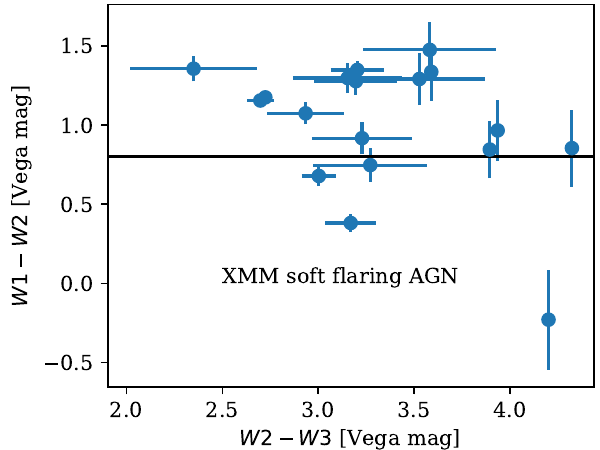}
\caption{Cross-match between soft flaring AGNs with \emph{WISE}. The horizontal lines $W1-W2>0.8$ mag indicate AGN candidates based on \emph{WISE} color.}
\label{fig:xray_wise}
\end{figure}

Next, we search for the IR counterpart of \xmm\ and \swift\ soft flaring AGNs in the \emph{WISE} \citep{wright2010} catalog. The dusty molecular torus that surrounds accreting SMBHs in AGNs absorbs the optical/UV light and then re-emits in IR wavelengths, with the emission typically peaking at around 15--60 $\mu$m \citep{mullaney2011}. We identified 18 of the 23 \xmm\ and all the \swift\ soft flaring AGNs with $\rm S/N>3$ in \emph{WISE} $W1$ and $W2$ band. Figure \ref{fig:xray_wise}  shows the \emph{WISE} color-color diagram of the \swift\ and \xmm\ flaring AGNs. A threshold of $W1-W2>0.8$ is used to identify the flaring sources as AGN via \emph{WISE} color at low redshift \citep{stern2012}. This method becomes less effective at high redshift as the \emph{WISE} color becomes progressively bluer with redshift in the range $2\lesssim z\lesssim 5$ \citep{assef2018}. Among the \emph{WISE} identified counterparts, 14/18 and 110/179 have $W1-W2>0.8$ and can be identified as AGNs in \xmm\ and \swift, respectively.

\begin{figure}[]
\centering
\includegraphics[width=\figsize\textwidth]{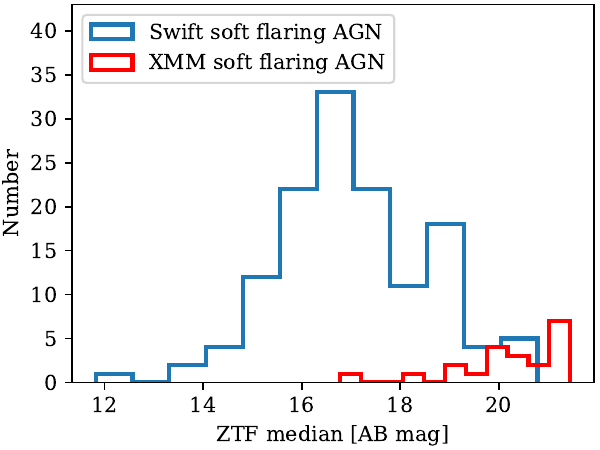}
\includegraphics[width=\figsize\textwidth]{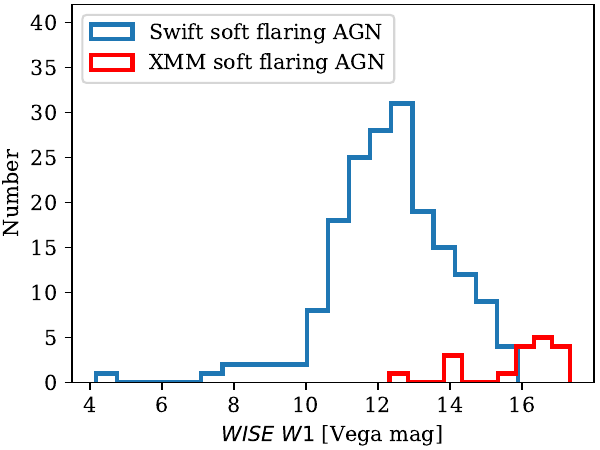}
\includegraphics[width=\figsize\textwidth]{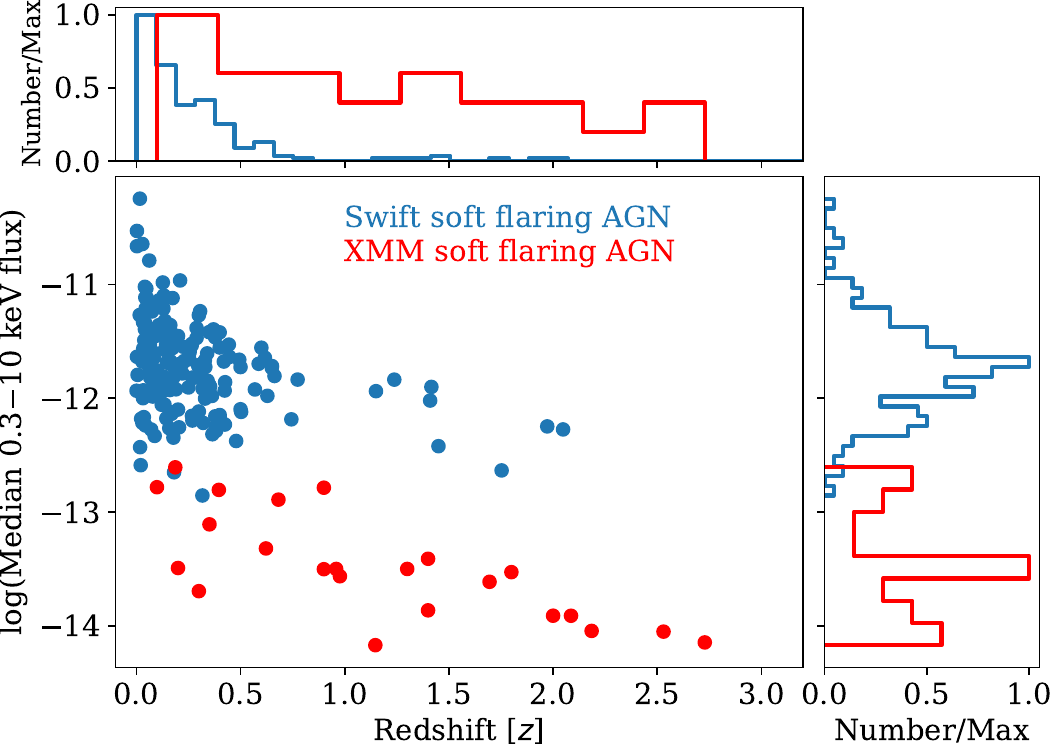}
\caption{Top panel: distribution of median mag for \swift\ and \xmm\ soft flaring AGNs for which we found a counterpart in ZTF. Middle panel: distribution of $W1$ values of the soft flaring AGNs for which we found a counterpart in \emph{WISE} catalog. Bottom panel: distribution of median 0.3--10 keV flux in units of erg s$^{-1}$ cm$^{-2}$ vs. redshift of the soft flaring AGNs.}
\label{fig:mean_ctr_redshift}
\end{figure}

In total, we were able to confirm 142\footnote{100 through ZTF, 110 through \emph{WISE}, including 68 in common.} (79\%) and 14\footnote{2 through ZTF, 14 through \emph{WISE}, including 2 in common.} (61\%) soft flaring sources as AGNs in \swift\ and \xmm, respectively, by combining optical variability from ZTF and IR color from \emph{WISE}. The AGN identification in the \xmm\ sample is slightly less robust than the \swift\ data set, primarily because the sources in the \xmm\ data set are much fainter than the ones in the \swift\ data set. The top and middle panels of Figure \ref{fig:mean_ctr_redshift} show the ZTF median AB mag and \emph{WISE} $W1$ Vega mag values of the flaring AGNs for which we find a counterpart, which indicate the \xmm\ sources are in the faint end of the distribution. The bottom panel of Figure \ref{fig:mean_ctr_redshift} shows the 0.3--10 keV flux and the redshift distribution of the flaring AGNs, which indicates that the majority of the \xmm\ sources are X-ray faint and located at a larger distance compared to sources in the \swift. Among the 23 \xmm\ soft flaring AGNs, 48\% are at $z>1$ with 0.3--10 keV flux below $3.9\times10^{-14}\rm\ erg\ s^{-1}\ cm^{-2}$ whereas in \swift\ among 179 soft flaring AGNs 84\% are at $z<0.5$ with 0.3--10 keV flux above $1.4\times10^{-13}\rm\ erg\ s^{-1}\ cm^{-2}$.

\section{Discussion}\label{sec:discussion}
\subsection{Origin of the extremely soft flares in AGNs}
\subsubsection{Soft X-ray excess}
Both the \xmm\ and \swift\ spectral analysis confirm that the AGNs tend to become softer during their flaring events, and this is likely due to the presence of an additional blackbody component. The blackbody component is not associated with the emission from the accretion disk in AGNs. For luminous black holes of $10^6$--$10^8\ M_{\odot}$, the emission from the thermal accretion disk peaks at the extreme ultraviolet (EUV), which is often referred to as the big blue bump \citep{shields1978,malkan1982}. The EUV band is unobservable due to Galactic photoelectric absorption. A large fraction of AGNs show a blackbody-like component in addition to the power-law-like continuum in the X-ray band, which is often termed as soft X-ray excess \citep{pravdo1981,boissay2016}. The blackbody component peaks below $\sim$1--2 keV band. The soft X-ray excess was thought to be the hard tail of UV blackbody emission from the accretion disk \citep{arnaud1985,leighly1999}. However, later this explanation was ruled out by the fact that for different black hole masses with varying accretion rates, the soft X-ray excess can be fitted by the same blackbody temperature, which is not expected for an accretion disk \citep{gierlinksi2004,porquet2004,bianchi2009}.

Two current popular models explain the soft X-ray excess: Comptonization of UV photons in a warm corona and ionized relativistically blurred reflection. In the first case, the UV photons from the accretion disk are Comptonized by an extended warm corona lying above the accretion disk that is optically thick ($\tau\sim10$--40) and cooler ($T\sim10^6$ K) than the hot corona, which is responsible for the primary X-ray emission continuum \citep{czerny1987,done2012,rozanska2015,petrucci2020,palit2024}. In the second case, ionized lines of N, O, Ne, Mg, and Si produced due to X-ray reflection from the accretion disk are relativistically blurred due to proximity to the BH, which leads to the excess emission below 2 keV \citep{crummy2006,fabian2009,garcia2010,walton2013}.

The exact origin of the soft X-ray excess is still debated. However, the contribution from the soft X-ray excess will make the overall X-ray continuum steeper, which leads to a lower HR value. We find that the soft X-ray excess varies systematically depending on the source state. The soft X-ray excess contributes significantly to the continuum for sources with lower values of HR ($<-0.8$). Furthermore, for a given HR bin, the soft X-ray excess tends to increase when the source is near or at the peak of the flare. The variation of soft X-ray excess emission is seen in individual sources as well. For example, \citet{kammoun2015} found for the case of IRAS 13224–3809, the excess flux above the continuum in the soft X-ray band linearly correlates with the continuum flux as $F_{\rm excess}\propto
F_{\rm primary}^{0.46}$, which suggests that as the source gets brighter, the soft X-ray excess becomes stronger. A similar trend was also seen in AGN Mrk 590 \citep{lawther2025}.

\subsubsection{Continuum steepening}
\begin{figure}[]
\centering
\includegraphics[width=\figsize\textwidth]{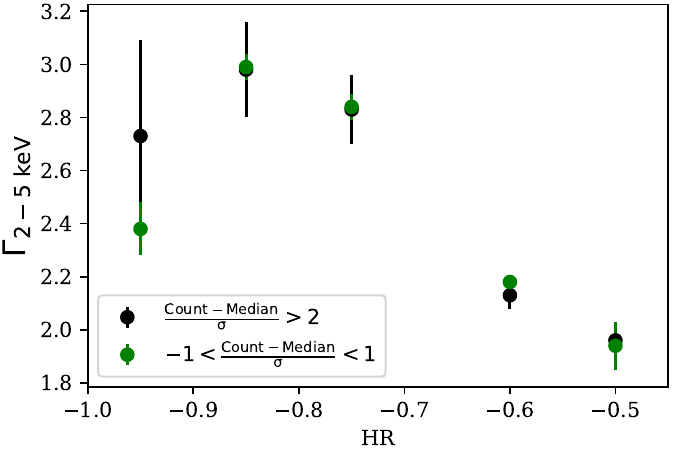}
\includegraphics[width=\figsize\textwidth]{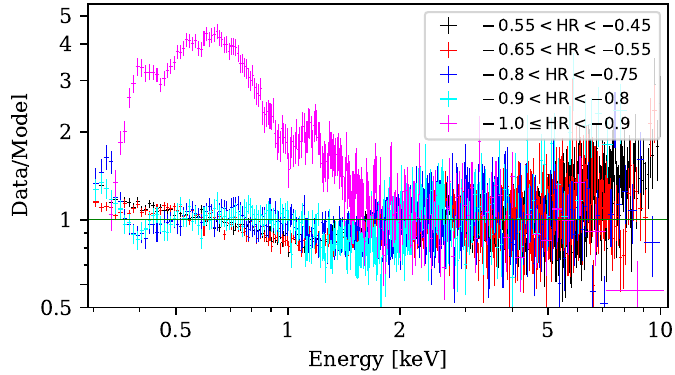}
\includegraphics[width=\figsize\textwidth]{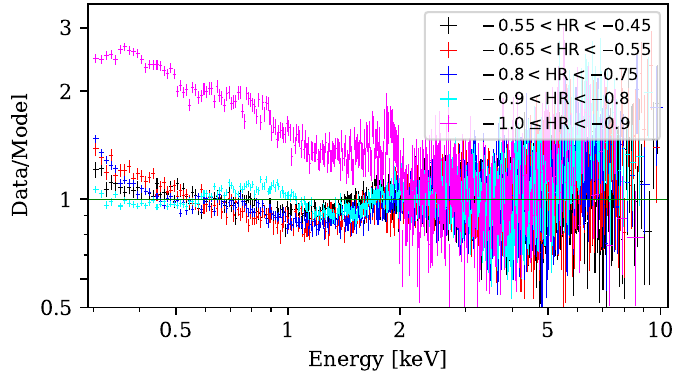}
\caption{Top panel: \swift\ AGN X-ray spectral photon index in 2--5 keV band for two different variability bins. Middle panel: excess at low energies when the 2--5 keV band best-fit model is extrapolated to lower energies for different HR values and the variability bin $\frac{\rm Count-Median}{\sigma}>2$. Bottom panel: same as the middle panel but for the variability bin $-1<\frac{\rm Count-Median}{\sigma}<1$.}
\label{fig:gamma_hr}
\end{figure}

The continuum emission or the power-law component may also vary with the source's variability. To test this, we fitted the \swift\ stacked spectra of different groups in the 2--5 keV band only and estimated the spectral photon index $\Gamma_{\rm 2-5\ keV}$ without including a blackbody component. If the soft X-ray excess component is solely responsible for the HR changes, then $\Gamma_{\rm 2-5\ keV}$ will be constant across the different HR groups. The top panel of Figure \ref{fig:gamma_hr} shows the $\Gamma_{\rm 2-5\ keV}$ for different HR values, which indicates in general the spectral slope becomes steeper for lower values of HR. We used 2--5 keV to estimate the photon index of the continuum X-ray emission, avoiding the soft X-ray excess that dominates below 2 keV and iron complex emission, which dominates in the 6--7 keV band (in cases of extreme relativistic effect, the red wing of Fe $K_{\alpha}$ can reach up to 5 keV). The middle and bottom panels of Figure \ref{fig:gamma_hr} show the strength of the soft X-ray excess when the 2--5 keV band best-fit model is extrapolated to lower energies. Indeed, the strength of the soft component increases for lower values of HR and higher values of $\frac{\rm Count-Median}{\sigma}$. However, the $\Gamma_{\rm 2-5\ keV}$ values for two different groups of $\frac{\rm Count-Median}{\sigma}$ are consistent with each other, which indicates no continuum variation for a given HR bin when the source is near the median value or in a flaring state. AGN continuum emission is known to exhibit X-ray spectral steepening as the Eddington accretion rate increases \citep{brightman2013,kubota2018}. Therefore, for a given $\frac{\rm Count-Median}{\sigma}$ bin, the linear correlation of HRs and $\Gamma_{\rm 2-5\ keV}$ in Figure \ref{fig:gamma_hr} is likely driven by a range of Eddington accretion rates across that sample. For example, higher values of HR or lower values of $\Gamma_{\rm 2-5\ keV}$ for a given $\frac{\rm Count-Median}{\sigma}$ bin might be associated with a lower level of Eddington accretion rate compared to lower values of HR. 

\begin{figure*}[]
\centering
\includegraphics[width=0.45\textwidth]{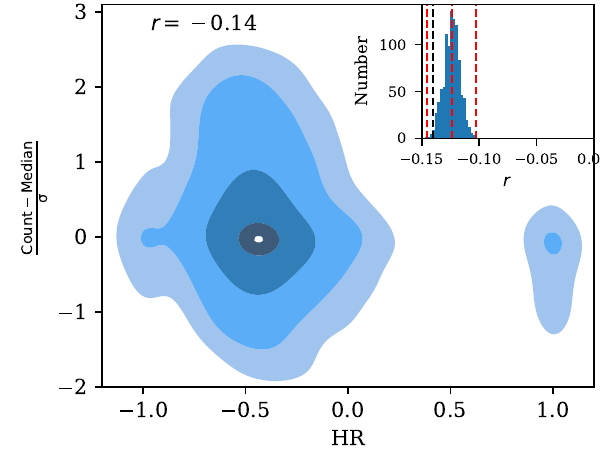}
\includegraphics[width=0.45\textwidth]{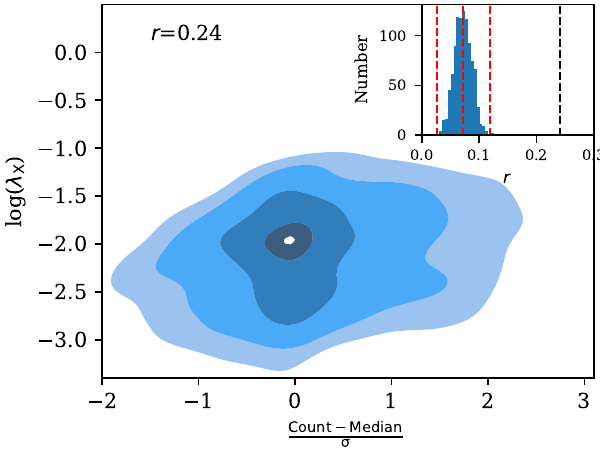}
\includegraphics[width=0.45\textwidth]{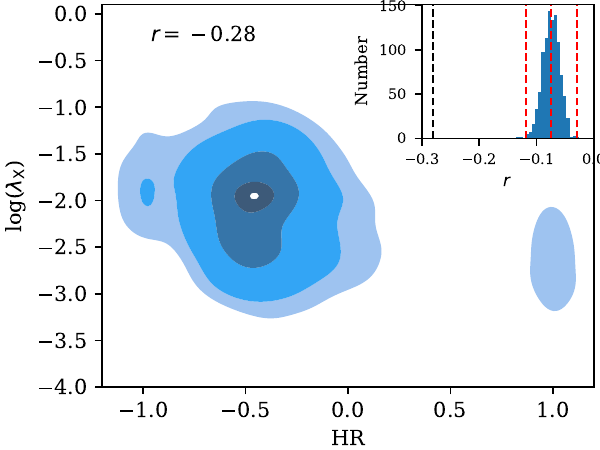}
\includegraphics[width=0.45\textwidth]{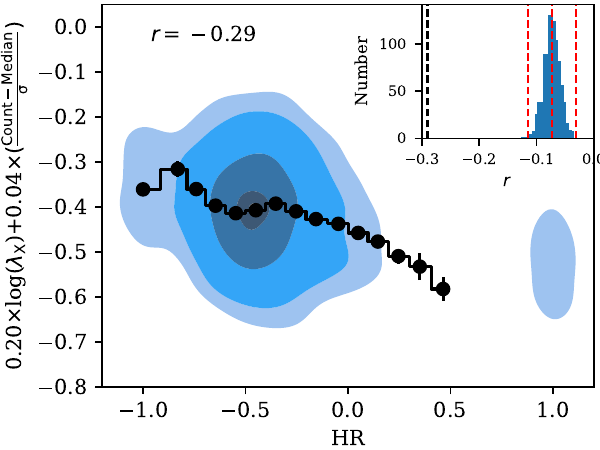}
\caption{Multidimensional correlation analysis of \swift\ AGN between the parameters HR, $\frac{\rm Count-Median}{\sigma}$, and $\lambda_{\rm X}$ with $r$ indicating the sample Pearson correlation coefficient without taking into account the measurement errors. The small inset shows the distribution of $r$ obtained from bootstrapping. A linear correlation between all combinations of HR and $\frac{\rm Count-Median}{\sigma}$, and log($\lambda_{\rm X}$) is found at a very high statistically significant level. However, HR depends more strongly on X-ray luminosity Eddington fraction $\lambda_{\rm X}$ than on flare strength $\frac{\rm Count-Median}{\sigma}$.}
\label{fig:swift_hr_var_mbh}
\end{figure*}

To test this hypothesis, we cross-match our \swift\ AGN sample to \citet{wu2022} catalog to obtain the masses of the SMBHs. The \citet{wu2022} catalog contains black hole masses of 750,414 quasars measured from Sloan Digital Sky Survey Data Release 16 optical spectra. Figure \ref{fig:swift_hr_var_mbh} shows the correlation analysis between the parameters HR, flare strength ($\frac{\rm Count-Median}{\sigma}$), and the X-ray luminosity Eddington fraction ($\lambda_{\rm X}$). We perform bootstrapping to quantify the effect of measurement error on correlation coefficient computation. Each pair of measurements is replaced by a simulated data point that is drawn from a Gaussian distribution with mean and standard deviation centered on the measurement and its uncertainties. We perform this test 1000 times and calculate the correlation coefficient. The small insets in Figure \ref{fig:swift_hr_var_mbh} show the distribution of the correlation coefficient obtained from the 1000 tests, with red dashed lines indicating the mean and $3\sigma$ boundaries. In all cases, the $3\sigma$ boundaries do not include zero correlation coefficient, indicating the statistical significance of the correlation at a $>3\sigma$ level. The black dashed lines in each small inset indicate the correlation coefficient obtained from considering only the measurements, and in cases that involve the BH masses, the correlation coefficient is significantly deviated from the bootstrapping mean due to large uncertainties in BH masses. Nonetheless, we still find a significant non-zero correlation between HR, $\frac{\rm Count-Median}{\sigma}$, and $\lambda_{\rm X}$. The top right panel of  Figure \ref{fig:swift_hr_var_mbh} indicates that the strength of the flare is highly correlated with the X-ray Eddington fraction. The top and bottom left panels indicate that the HR depends more strongly on $\lambda_{\rm X}$ than $\frac{\rm Count-Median}{\sigma}$, suggesting the accretion rate primarily determines the HR values; however, the flares will help to push the source to further lower HR values. A linear regression analysis between the three parameters indicates $\rm HR=(0.19\pm0.01)\times log(\lambda_{\rm X})+(0.035\pm0.006)\times(\frac{\rm Count-Median}{\sigma})-0.73\pm0.02$, suggesting X-ray spectral index $\Gamma_{2-5\rm\ keV}$ is primarily driven by the X-ray Eddington fraction $\lambda_{\rm X}$.

\begin{figure*}[]
\centering
\includegraphics[width=\figsizeee\textwidth]{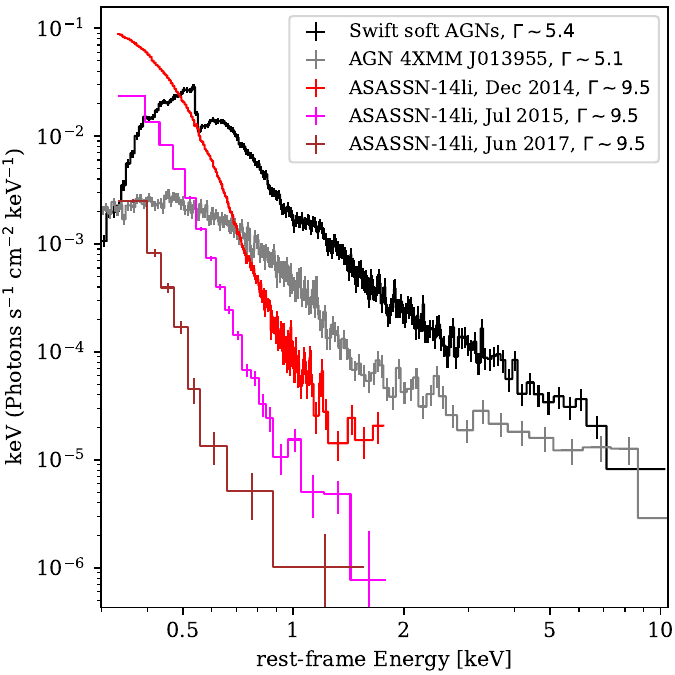}
\includegraphics[width=\figsizeee\textwidth]{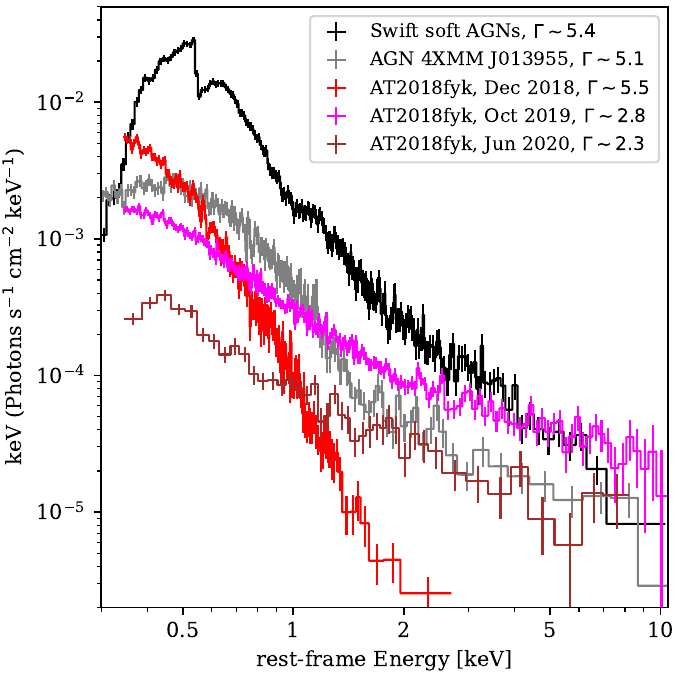}
\includegraphics[width=\figsizeee\textwidth]{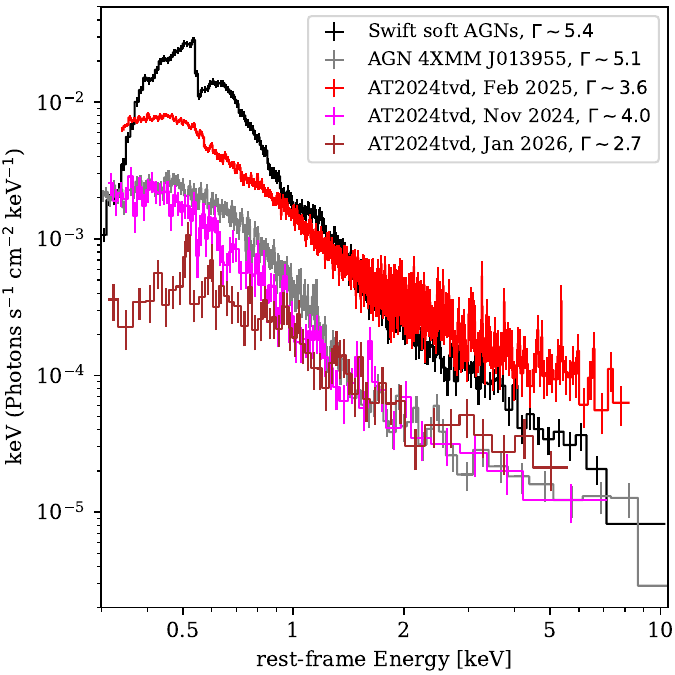}
\caption{Comparison of 0.3--10 keV X-ray spectral shape of softest AGN in \xmm\ 4XMM J013955.7+061922 and \swift\ soft AGNs ($\frac{\rm Count-Median}{\sigma}>2$ and $\rm -1.0\leq HR<-0.9$) with three well-known TDEs: ASASSN-14li (left panel), AT2018fyk (middle panel) and AT2024tvd (right panel). The spectral shape is obtained by fitting an absorbed blackbody plus power-law model, whereas $\Gamma$ indicates the broad-band spectral index obtained from an absorbed power-law model. A remarkable similarity can be seen between X-ray spectra of soft AGNs and post-peak spectra of AT2018fyk and AT2024tvd. The spectra of ASASSN-14li, AT2018fyk, and AT2024tvd in the Feb 2025 are taken from \xmm\ archive. The spectra of AT2024tvd in Nov 2024 and Jan 2026 are obtained by stacking of \swift\ observations.}
\label{fig:tde_agn_spec}
\end{figure*}

\subsection{X-ray spectra comparison of soft AGN flares and some well-known TDEs}
Next, we compare the shape of X-ray spectra of TDEs with flaring soft AGNs. For this comparison, we choose three well-known TDEs: ASASSN-14li, AT2018fyk, and AT2024tvd, which show very different X-ray spectral evolution. The spectra of ASASSN-14li are dominated by a blackbody component at all times. AT2018fyk only shows the blackbody component during the peak; however, after the peak, an extra powerlaw component emerges. Lastly, AT2024tvd never shows a single blackbody-dominated state whose spectra are best described by a blackbody plus power-law at all times. Figure \ref{fig:tde_agn_spec} shows the comparison of the X-ray spectral shape of TDEs with soft flaring AGNs. The TDEs show a diverse set of X-ray spectra, and only in the cases where the emission is completely dominated by the thermal emission can they be distinguished from the soft flaring AGNs. This includes ASASSN-14li, which does not show any significant spectral evolution, and AT2018fyk during the peak. On the other hand, the post-peak spectra of AT2018fyk and AT2024tvd show a remarkable similarity with the spectra of soft flaring AGNs. We also mention the overall spectral photon index $\Gamma$ by fitting an absorbed power-law model. The spectral index can only be used to separate cases like ASASSN-14li, which has a $\Gamma\sim9.5$ much larger than soft flaring AGNs with $\Gamma\sim5$. Whereas for AT2018fyk and AT2024tvd, the $\Gamma$ values are comparable or even lower than those of soft flaring AGNs. This indicates that separating TDEs and soft flares from AGNs solely based on X-ray HR or spectral index $\Gamma$ can be extremely challenging, especially if there is a significant emission above 2 keV in TDE spectra.

\citet{auchettl2018} compared the X-ray emission properties of four bright TDEs and a sample of AGNs. They found that highly variable AGN can produce flares of a similar magnitude that are seen in X-ray TDEs. Furthermore, the peak of an AGN flare tends to have quite soft X-ray emission, similar to our findings. Recently \citet{guolo2024} compared optically selected X-ray bright TDEs and 70-month time-averaged X-ray spectral properties of \emph{Swift}-BAT type I and II AGNs \citep{ricci2017}. In their sample, 85\% of TDEs have $\rm HR<-0.85$ whereas type I and II AGNs are concentrated at $\rm -0.7\leq HR\leq0.0$, and $\rm 0.0\leq HR\leq1.0$, respectively, finding that X-ray emission from TDEs is much softer with a clear separation from AGNs. We find similar values when averaged over all AGNs over all times in the sample. Similarly, \citet{chakraborty2026} find that while TDEs have a softer spectral index than AGN in general, TDEs can have coronae, and can have spectral properties that overlap with the population of AGN. In the section below, we consider how the significant overlap in spectral properties of TDEs and AGN can contribute to AGN contamination in TDE searches, or cases where TDEs could be missed due to harder X-ray spectra. In this work, we look specifically at AGNs during their flaring states, which show systematically softer X-ray spectra exactly when they are most likely to mimic TDEs. Our findings are thus different than those of \citet{guolo2024}, primarily due to the fact that AGNs can have a wide range of X-ray spectral properties, which may not be revealed when looking at time-averaged X-ray emission. 

\subsection{Rate of soft flares from AGNs}
We estimate the rate of soft flares in AGN from the \xmm\ and \swift\ sample. These rate estimates represent a lower limit as \xmm\ and \swift\ are not well sampled on a regular cadence, which increases the chances of missing a flare. Therefore, the actual rate of soft flares from AGN may be much higher than our estimates. The rates of AGN flares will depend on the strength and duration assumptions used to identify flares, as we consider further in \citet{mondal2025}.

We detected 23 (2.5\%) flares out of 920 in \xmm\ and 179 (4.4\%) flares out of 4089 in \swift\ AGN sample, that have $\frac{\rm Count-Median}{\sigma}>2$ and peak $\rm HR<-0.75$. We estimate the total AGN years for the \xmm\ and \swift\ AGN sample in the rest-frame. The \xmm\ and \swift\ cover a total of $1.94\times10^3$ and $1.95\times10^4$ AGN years, respectively. The rate of soft flares from AGNs in \xmm\ and \swift\ is $1.2\times10^{-2}\rm\ AGN^{-1}\ yr^{-1}$ and $9.2\times10^{-3}\rm\ AGN^{-1}\ yr^{-1}$, respectively. To compare the estimated rate with TDEs, one has to convert flares per AGN to per galaxy. The fraction of galaxies hosting an AGN depends on various factors such as stellar mass, redshift, and galaxy type. Typically, the AGN fraction can range anywhere between 9.3\% at $0.5<z<1.5$ up to $21.2\%$ at $1.5<z<2.5$ \citep{wang2017}. Therefore, the rate of AGN flares with $\frac{\rm Count-Median}{\sigma}>2$ and peak $\rm HR<-0.75$ in galaxy terms is $(1.1-2.5)\times10^{-3}\rm\ galaxy^{-1}\ yr^{-1}$ and $(0.9-2.0)\times10^{-3}\rm\ galaxy^{-1}\ yr^{-1}$ in \xmm\ and \swift, respectively, for considering an AGN fraction of 9.3--21.2\%. 

The current estimate of the TDE rate from optical observations is $3.2\times10^{-5}\rm\ galaxy^{-1}\ yr^{-1}$ \citep{yao2023}. The rate of soft AGN flares with strength $>2\sigma$ from the median is 34--147 times higher than the current rate of TDEs from optical surveys. 

\begin{figure}[]
\centering
\includegraphics[width=\figsize\textwidth]{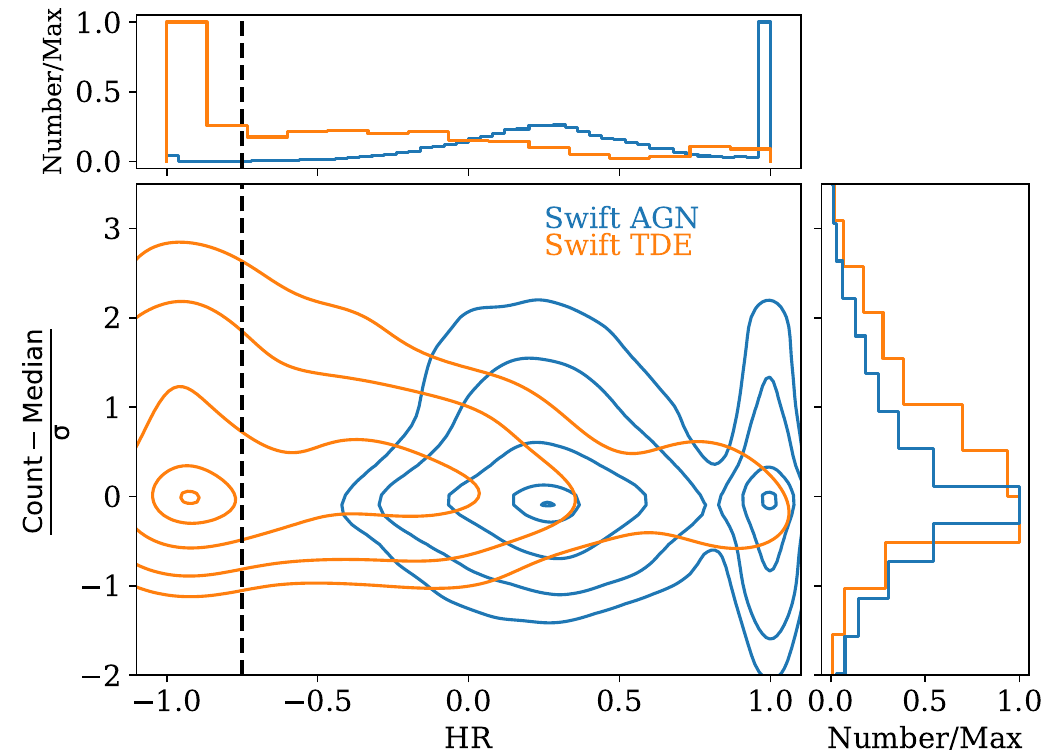}
\caption{Comparison of TDE and AGN HR constructed from counts in 0.3--1 and 1--10 keV bands. A threshold of $\rm HR<-0.75$ (black dashed line) can be used to clearly separate the softest TDEs from flaring soft AGNs, which was previously difficult in Figure \ref{fig:hr_agn_tde}. However, using this threshold, a significant number of TDEs with hard X-ray emission will be missed.}
\label{fig:hr_agn_tde1}
\end{figure}

We also compute the rate of flares with maximum flux change (peak/min) by a factor of 20 or more happening on a time scale greater than two years with peak $\rm HR<-0.75$. There are only 1 and 7 such AGNs in \xmm\ and \swift, respectively. The rate of such AGN flares is $(4.7-11.0)\times10^{-5}\rm\ galaxy^{-1}\ yr^{-1}$ in \xmm\ and $(3.1-7.1)\times10^{-5}\rm\ galaxy^{-1}\ yr^{-1}$ \swift, considering an AGN fraction of 9.3--21.2\%. The rate of soft flare with maximum flux difference by a factor of 20 or more with peak $\rm HR<-0.75$ is comparable to the current optical TDE rate. This highlights that TDE searches using HR cuts may still be contaminated by flaring events from AGNs. However, the soft flares from AGNs contaminating X-ray TDE searches can be reduced by selecting a threshold of 20 or more in maximum flux change happening on a timescale of more than two years, as well as by using multi-wavelength AGN indicators as discussed in \S\ref{sec:counterparts}.

We test if a different definition of HR can be used to separate TDEs with blackbody-dominated spectra from soft flaring AGNs. We constructed the new HR based on the counts in 0.3--1 and 1--10 keV bands. Figure \ref{fig:hr_agn_tde1} shows the comparison of TDEs and AGNs HR using the above bands. As the soft flaring AGN always have significant emission at hard X-ray, therefore, using the above definition of HR will push the soft flaring AGN towards a more positive value, whereas the soft TDEs will remain relatively unchanged, as their entire emission is contained below $\sim$1--2 keV. Indeed, a $\rm HR<-0.75$ cut can clearly separate the softest TDEs from flaring AGN, which was previously difficult in Figure \ref{fig:hr_agn_tde}, but it also becomes indistinguishable between TDEs with significant hard emission from AGNs.

\begin{figure}[]
\centering
\includegraphics[width=\figsize\textwidth]{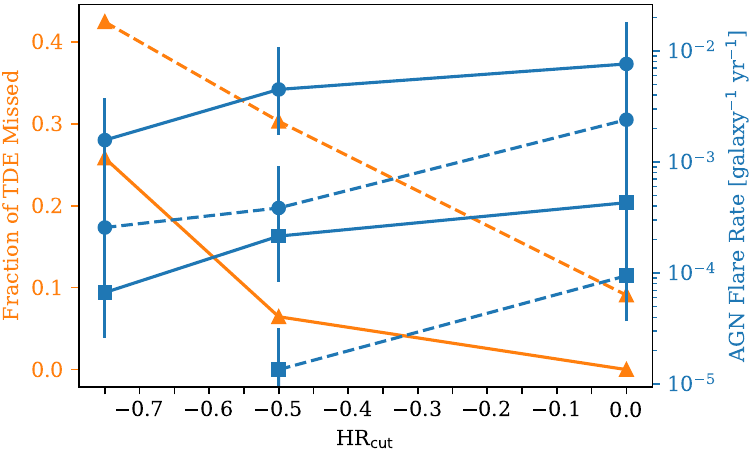}
\caption{Points connected with solid line indicate HR computed from 0.3--2 and 2--10 keV band, whereas points connected with dashed line indicate HR computed from 0.3--1 and 1--10 keV band. The left Y-axis indicates the fraction of TDE missed for three HR cuts: $<-0.75$, $<-0.5$, and $<0.0$. Right Y-axis indicates AGN flare rate (circle: $\frac{\rm Count-Median}{\sigma}>2$; square: $\frac{\rm peak}{\rm min}>20\times$, and $|t_{\rm peak}-t_{\rm min}|>2\ \rm yr$) for various HR cuts.}
\label{fig:tde_missed}
\end{figure}

Figure \ref{fig:tde_missed} shows the effect of different HR cuts that can be used to separate TDEs and their effect on AGN flare rate. Choosing the HR in 0.3--1 and 1--10 keV bands (dashed lines) leads to the lowest rate of AGN flare. However, this also leads to the largest number of TDEs being missed, 42\% for $\rm HR<-0.75$ and 30\% for $\rm HR<-0.5$. On the contrary, HR defined using 0.3--2 and 2--10 keV bands leads to the lowest number of TDE being missed, around 26\% for $\rm HR<-0.75$ and 6\% for $\rm HR<-0.5$. In general (both solid and dashed lines), increasing the HR cut lowers the fraction of TDE being missed. However, this also leads to more overlap with soft AGNs, which increases the soft AGN flare rate.

\section{Conclusions}\label{sec:conclusions}
We compare the X-ray emission properties of TDEs and AGNs to determine if there is a significant overlap or if they can be separated based on HR. We search for AGNs that have soft flare-like events mimicking TDEs in \xmm\ and \swift\ catalogs. The conclusions of this study are as follows.

\begin{itemize}
    \item AGN tend to become softer during their flaring events and can have similar spectral properties as TDEs. The softening is primarily driven by the steepening of the continuum due to an increase in accretion rate as well as an increase in the soft X-ray excess strength.
    \item We detected flares with strength $>2\sigma$ from the median count rate that have peak $\rm HR<-0.75$ in 2.5\% and 4.4\% of the AGN samples of \xmm\ and \swift, respectively. The rate of soft flares with strength $>2\sigma$ from the median is at least 34--147 times higher than current TDE rates, indicating future X-ray TDE searches relying only on rise or power-law-like decay will be highly contaminated by small magnitude soft flares from AGNs. However, choosing a threshold in maximum flux change by $>$ 20 times on a longer time scale of two years or more with peak $\rm HR<-0.75$ can reduce this contamination, as the rate of such flares in AGNs is comparable or lower than current TDE rates. 
    \item These flaring AGN can be identified with prior optical or IR observations, but the effectiveness of this identification highly depends on the distance and luminosity of the source. For example, the \swift\ sample traces the local bright AGN population in which $\sim79\%$ flaring sources can be identified as AGN; on the contrary, the \xmm\ sample is much fainter at larger distances and only $\sim61\%$ can be identified as AGN through optical/IR properties.
\end{itemize}

In our previous work of \citet{mondal2025}, based on the variability in the light curves only, we have highlighted that the many TDEs discovered using sparsely sampled X-ray observations have a similar amount of maximum variation as AGN flares. Building upon the previous work in this paper, we find that AGNs and TDEs can have not only a similar magnitude of variation but also very similar X-ray spectral properties. Therefore, identifying new TDEs in currently ongoing X-ray surveys such as eROSITA and Einstein Probe should be done more cautiously, such as utilizing multi-wavelength follow-up or considering a longer baseline with deeper exposure for X-ray observations that is long enough to distinguish between power law decay or stochastic variability.

%% Also note that the akcnowlodgment environment does not support long amounts of text. If you have a lot of people and institutions to acknowledge, do not use this command. Instead, create a new \section{Acknowledgments}.
\begin{acknowledgments}
SM and KDF acknowledge support from grant NASA ADAP 80NSSC24K0666. SM acknowledges support from NASA \nustar\ data analysis funding 1729326. J.T.H acknowledges support provided by NASA through the NASA Hubble Fellowship grant HST-HF2-51577.001-A awarded by the Space Telescope Science Institute, which is operated by the Association of Universities for Research in Astronomy, Incorporated, under NASA contract NAS5-26555. 

We thank the anonymous referee for useful feedback, which has improved this manuscript. SM thanks Shi-Jiang Chen for help with \texttt{Xstack} and feedback on the manuscript.
\end{acknowledgments}


\vspace{5mm}
\facilities{Swift, XMM, TNS, and \emph{WISE} }


\software{python \citep{vanrossum2009}, jupyter \citep{kluyver2016}, astropy \citep{2013A&A...558A..33A,2018AJ....156..123A,2022ApJ...935..167A}, numpy \citep{vanderwalt2011,harris2020}, matplotlib \citep{hunter2007}}


\appendix

\section{Variation of X-ray spectral components based on $\lambda_{\rm X}$ and HR}
\begin{figure}[]
\centering
\includegraphics[width=\figsize\textwidth]{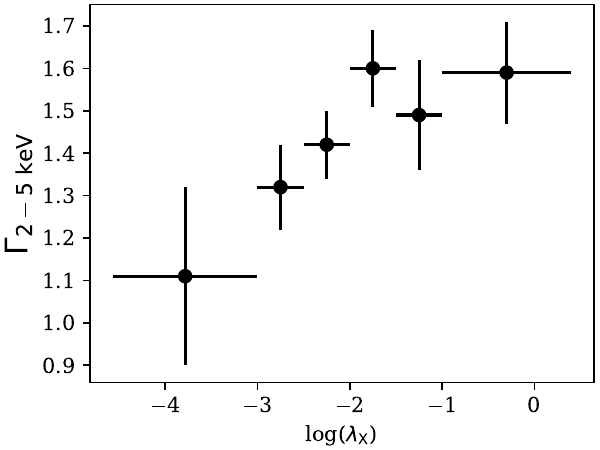}
\caption{The linear correlation between X-ray spectral photon index $\Gamma_{\rm 2-5\ keV}$ and $\lambda_{\rm X}$, which indicates the continuum emission gets steeper, possibly due to a higher accretion rate. }
\label{fig:swift_stack_gamma_lambdax}
\end{figure}

\begin{figure*}[]
\centering
\includegraphics[width=\figsizeee\textwidth]{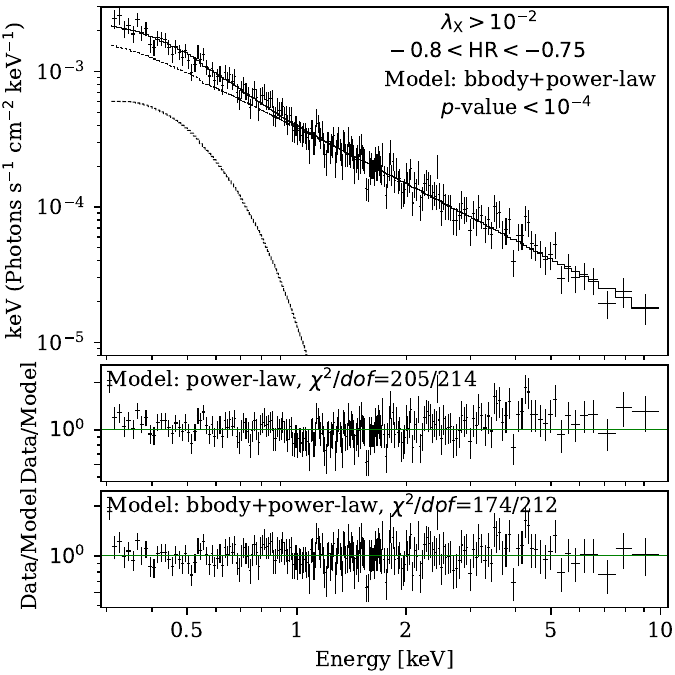}
\includegraphics[width=\figsizeee\textwidth]{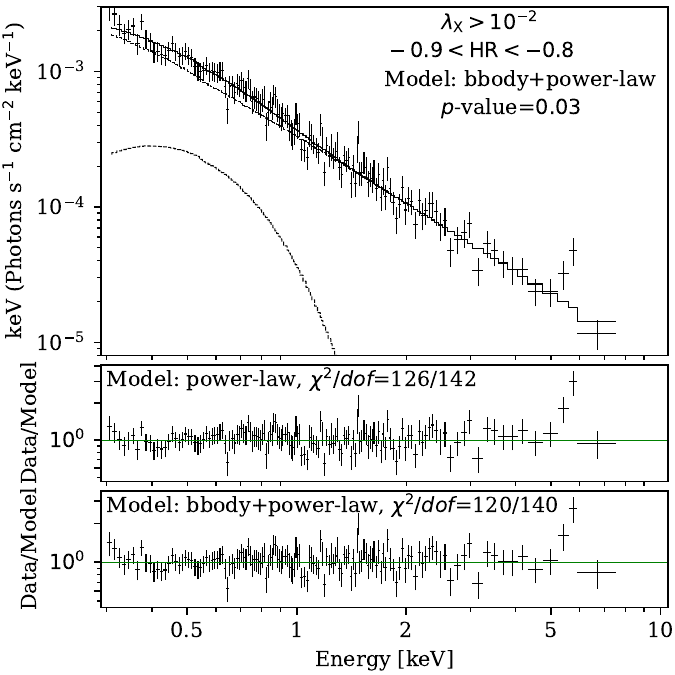}
\includegraphics[width=\figsizeee\textwidth]{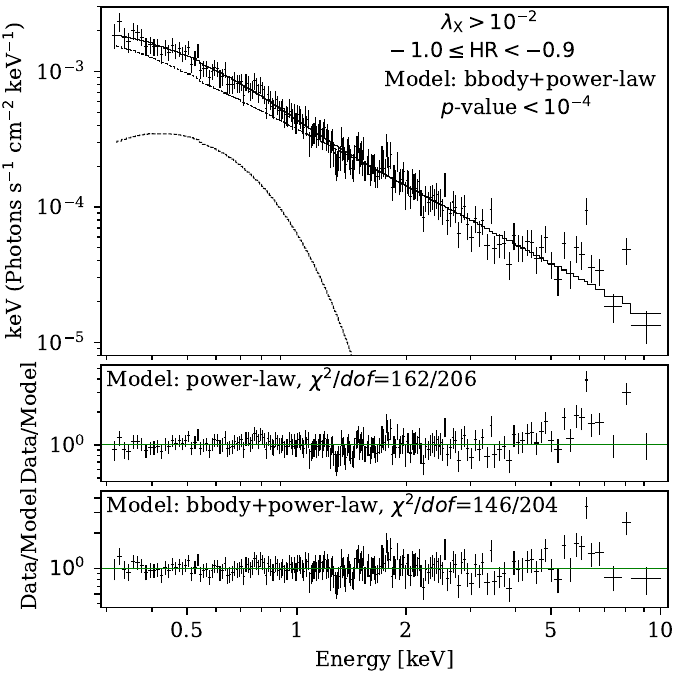}
\includegraphics[width=\figsizeee\textwidth]{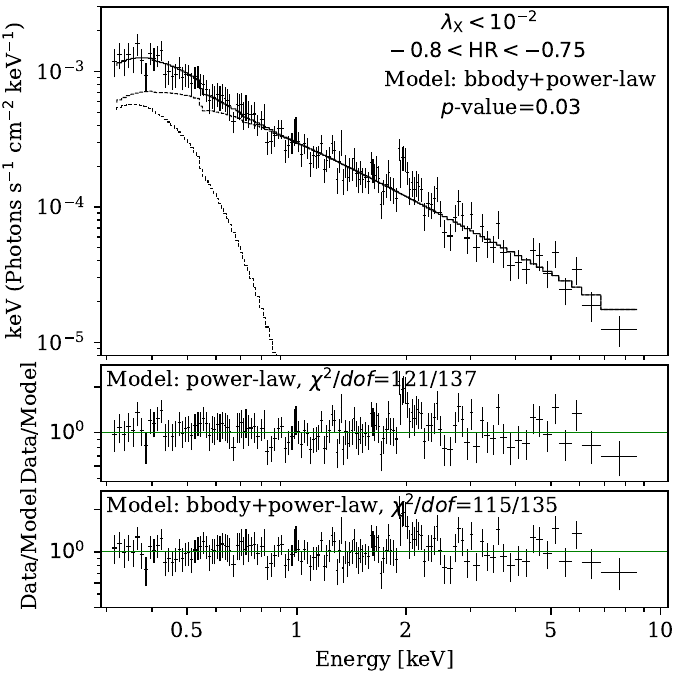}
\includegraphics[width=\figsizeee\textwidth]{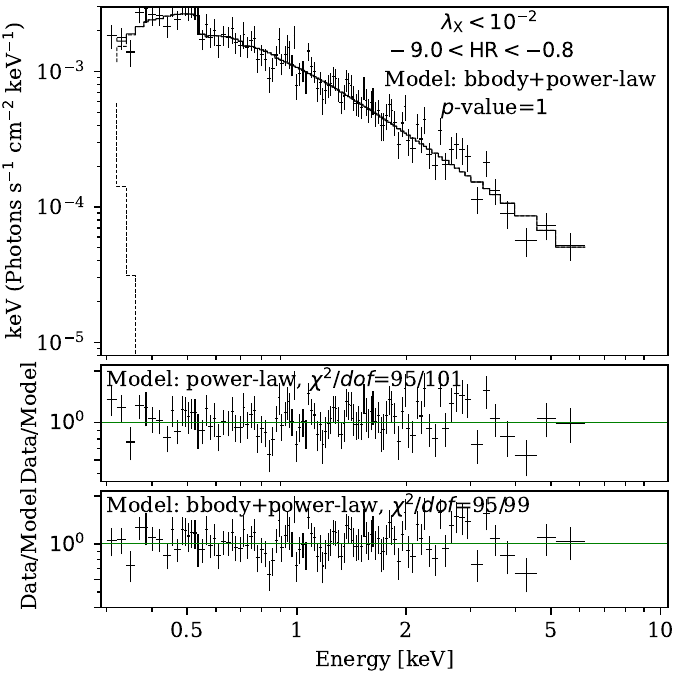}
\includegraphics[width=\figsizeee\textwidth]{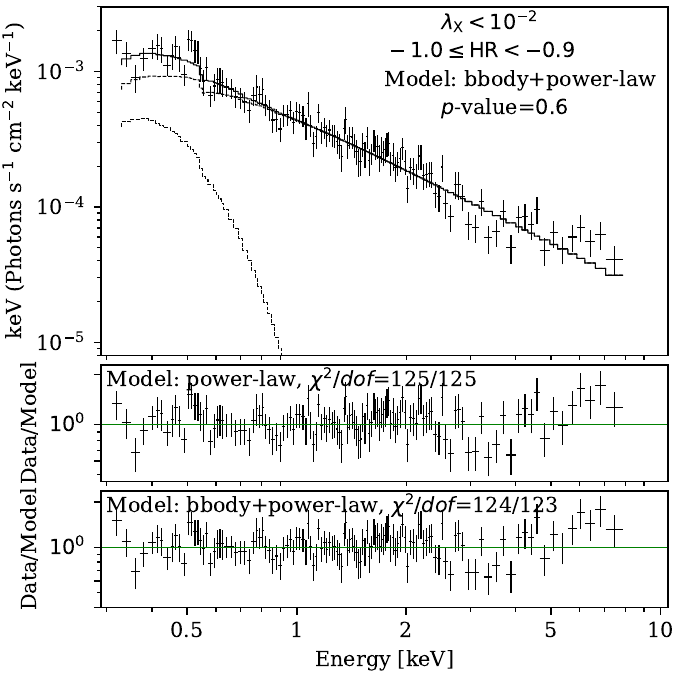}
\caption{\swift\ stack spectra of soft AGNs for different HR and $\lambda_{\rm X}>10^{-2}$ (top panels) and $\lambda_{\rm X}<10^{-2}$ (bottom panels) groups. Only for the top panels ($\lambda_{\rm X}>10^{-2}$), adding a blackbody component provides a significant improvement in overall fit statistics.}
\label{fig:swift_stack_ledd}
\end{figure*}

For the sample of \swift\ and \citet{wu2022} cross-matched AGN, we performed stacked spectral analysis based on various $\lambda_{\rm X}$ and HR groups. Figure \ref{fig:swift_stack_gamma_lambdax} shows the X-ray spectral photon index $\Gamma_{\rm 2-5\ keV}$ for the stacked spectra of various $\lambda_{\rm X}$ groups. A linear correlation is present between $\Gamma_{\rm 2-5\ keV}$ and $\lambda_{\rm X}$, indicating the continuum emission is getting steeper, possibly due to a higher accretion rate that further contributes to lowering HR values in the 0.2--2 and 2--10 keV bands besides the soft X-ray excess. The $\Gamma_{\rm 2-5\ keV}$ in Figure \ref{fig:swift_stack_gamma_lambdax} are significantly harder than Figure \ref{fig:gamma_hr}. This is primarily due to the fact that Figure \ref{fig:gamma_hr} contains AGNs with $\rm HR<-0.5$ whereas in Figure \ref{fig:swift_stack_gamma_lambdax} we select all unobscured AGNs with $\rm HR<0.5$ to increase the sample size and the number of bins. \citet{sobolewska2009} obtained a similar positive correlation between $\Gamma$ and mass accretion rate from a sample of AGN observed by \textit{RXTE}. However, in their case, the average $\Gamma$ is around 1.8, whereas in Figure \ref{fig:swift_stack_gamma_lambdax} all values are below 1.8. This is primarily because we measure $\Gamma_{\rm 2-5\ keV}$ to avoid the soft X-ray excess, which dominated below 2 keV as well as iron complex emission in the 6--7 keV band, whereas \citet{sobolewska2009} used the entire \textit{RXTE} band 3--20 keV. Furthermore, Figure \ref{fig:swift_stack_ledd} shows the spectra for $\lambda_{\rm X}>10^{-2}$ and $<10^{-2}$ for three HR groups: $\rm -0.8<HR<-0.75$, $\rm -0.9<HR<-0.8$, $\rm -1.0\leq HR<-0.9$. The stacked spectra are fitted with an absorbed power-law and blackbody plus power-law model. The blackbody plus power-law model is strongly favored for the $\lambda_{\rm X}>10^{-2}$, which is likely tracing more flaring events as $\lambda_{\rm X}$ and $\frac{\rm Count-Median}{\sigma}$ show strong correlation in Figure \ref{fig:swift_hr_var_mbh}.

\startlongtable
\begin{deluxetable*}{c c c c c c}
\label{tab:tde_list}
\tablecaption{Details of the number of observations available for the TDE sample in \swift\ and \xmm}
\tablehead{
   Name & RA & DEC & $z$ & $N_{\rm XRT}$ & $N_{\rm XMM}$
}
\startdata
2MASXJ01190869-3411305  &  19.7833  &  -34.1917  &  0.018  &  96 & 2\\ \hline
3XMMJ150052.0+015452  &  225.2167  &  1.9149  &  0.145  &  2 & 6\\ \hline
ASASSN-14li  &  192.0625  &  17.7740  &  0.021  &  101 \\ \hline
ASASSN-15oi  &  309.7875  &  -30.7556  &  0.048  &  23 & 2\\ \hline
AT2018fyk  &  342.5672  &  -44.8649  &  0.059  &  168 \\ \hline
AT2019azh  &  123.3208  &  22.6483  &  0.022  &  37 \\ \hline
AT2019dsg  &  314.2625  &  14.2056  &  0.051  &  7 \\ \hline
AT2020ocn  &  208.4750  &  53.9972  &  0.070  &  66 \\ \hline
CSS100217  &  157.3042  &  40.7055  &  0.147  &  14 \\ \hline
F01004-2237  &  15.7083  &  -22.3658  &  0.118  &  5 \\ \hline
IGRJ17361-4441  &  264.0708  &  -44.7356  &  0.000  &  38 & 4\\ \hline
NGC247  &  11.7875  &  -20.7603  &  0.001  &  17 & 3\\ \hline
OGLE16aaa  &  16.8375  &  -64.2725  &  0.166  &  2 & 2\\ \hline
SDSSJ1201+30  &  180.4000  &  30.0515  &  0.146  &  2 & 3\\ \hline
SRGEJ071310.6+725627  &  108.2938  &  72.9408  &  0.104  &  4 \\ \hline
SwiftJ1112-82  &  167.9500  &  -82.6458  &  0.890  &  9 \\ \hline
XMMSL1J024916.6-041244  &  42.3208  &  -4.2144  &  0.019  &  2 \\ \hline
XMMSL1J061927.1-655311  &  94.8625  &  -65.8853  &  0.073  &  16 & 2\\ \hline
XMMSL1J131952.3+225958  &  97.6917  &  -60.5194  &  0.014  &  4 \\ \hline
XMMSL2J140446.9-251135  &  211.1958  &  -25.1931  &  0.043  &  13 \\ \hline
XMMSL2J144605.0+685735  &  221.5208  &  68.9586  &  0.029  &  17 & 2\\ \hline
TDE2025cyj  &  141.3326  &  23.5026  &  0.144  &  7 \\ \hline
TDE2025chm  &  115.2460  &  16.5316  &  0.101  &  40 \\ \hline
TDE2024aepd  &  231.4137  &  17.6807  &  0.084  &  8 \\ \hline
TDE2024admt  &  158.4954  &  14.6881  &  0.111  &  7 \\ \hline
TDE2024ymu  &  1.7190  &  -17.3368  &  0.120  &  2 \\ \hline
TDE2024tvd  &  257.6774  &  28.8375  &  0.045  &  49 \\ \hline
TDE2024rny  &  344.3700  &  39.4290  &  0.105  &  3 \\ \hline
TDE2024qab  &  273.5258  &  36.2499  &  0.140  &  2 \\ \hline
TDE2024mvz  &  355.1158  &  -7.0144  &  0.080  &  16 \\ \hline
TDE2024lhc  &  252.7258  &  32.8767  &  0.204  &  34 \\ \hline
TDE2023mhs  &  205.8153  &  19.2503  &  0.048  &  6 \\ \hline
TDE2023lli  &  344.4145  &  40.5444  &  0.036  &  99 \\ \hline
TDE2023cvb  &  288.6070  &  41.6692  &  0.071  &  21 \\ \hline
TDE2022upj  &  5.9869  &  -14.4231  &  0.054  &  33 \\ \hline
TDE2022lri  &  35.0334  &  -22.7209  &  0.033  &  82 \\ \hline
TDE2022gri  &  109.5866  &  33.9949  &  0.028  &  12 \\ \hline
TDE2022fpx  &  232.7654  &  53.4053  &  0.073  &  22 \\ \hline
TDE2022exr  &  262.4605  &  25.8422  &  0.096  &  44 \\ \hline
TDE2022dyt  &  150.0334  &  26.4607  &  0.072  &  8 \\ \hline
TDE2022dsb  &  235.5906  &  -22.6706  &  0.023  &  41 \\ \hline
TDE2022czy  &  185.5047  &  16.9956  &  0.109  &  17 \\ \hline
TDE2022arb  &  156.0430  &  -0.8230  &  0.062  &  5 \\ \hline
TDE2021uqv  &  8.1662  &  22.5489  &  0.106  &  2 \\ \hline
TDE2021ehb  &  46.9492  &  40.3113  &  0.018  &  82 \\ \hline
TDE2020afhd  &  48.3987  &  -2.1518  &  0.027  &  89 \\ \hline
TDE2020nov  &  254.5540  &  2.1175  &  0.084  &  21 \\ \hline
TDE2019vcb  &  189.7349  &  33.1659  &  0.089  &  2 \\ \hline
TDE2019teq  &  284.7729  &  47.5182  &  0.088  &  30 \\ \hline
TDE2019qiz  &  71.6578  &  -10.2264  &  0.015  &  15 \\ \hline
TDE2019ehz  &  212.4245  &  55.4911  &  0.074  &  7 \\ \hline
2XMMJ123103.2+110648  &  187.7625  &  11.1135  &  0.119  &&  6 \\ \hline
2XMMiJ184725.1-631724  &  281.8542  &  -63.2903  &  0.035  &&  3 \\ \hline
3XMMJ152130.7+074916  &  230.3792  &  7.8213  &  0.179  &&  2 \\ \hline
NGC3599  &  168.8625  &  18.1104  &  0.003  &&  4 \\ \hline
\enddata
\end{deluxetable*}

\bibliography{sample631}{}
\bibliographystyle{aasjournal}



\end{document}